\documentclass[journal]{IEEEtran}
\usepackage[T1]{fontenc}

\usepackage{cite}
\usepackage{amsthm}
\usepackage{amsmath,graphicx}
\usepackage{amssymb}
\usepackage{amsfonts}
\usepackage{graphicx,subfigure}
\usepackage{fancyhdr}  %
\usepackage{cases}
\usepackage{extarrows}
\usepackage{algorithm}
\usepackage{algorithmic}
\usepackage{multirow,tabularx}
\usepackage{mathtools}
\usepackage{caption}
\usepackage{bm}
\usepackage{graphicx}
\usepackage{epstopdf}
\usepackage{cuted}
\usepackage[table]{xcolor}

\DeclareMathOperator*{\argmin}{arg\,min}

\expandafter\def\expandafter\normalsize\expandafter{%
	\normalsize%
	\setlength\abovedisplayskip{4.2pt}%
	\setlength\belowdisplayskip{4.2pt}%
	\setlength\abovedisplayshortskip{2pt}%
	\setlength\belowdisplayshortskip{2pt}%
}

\theoremstyle{definition}

\newtheorem{lemma}{Lemma}
\newtheorem{remark}{Remark}

\usepackage{hyperref}
\hypersetup{colorlinks=true,
	linkcolor=blue,
	citecolor=blue,      
	urlcolor=black,
}

\makeatletter
\renewcommand{\maketag@@@}[1]{\hbox{\m@th\normalsize\normalfont#1}}%
\makeatother

\begin{document}
	
	\title{Dual-Scale Antenna Deployment for \\ Pinching Antenna Systems}

	\author{Xu Gan,~\IEEEmembership{Member, IEEE}, Zhaolin Wang,~\IEEEmembership{Member, IEEE}, Yuanwei Liu,~\IEEEmembership{Fellow,~IEEE}
		\thanks{The authors are with the Department of Electrical and Computer Engineering, The University of Hong Kong, Hong Kong (e-mail: \{eee.ganxu, zhaolin.wang, yuanwei\}@hku.hk).}\vspace{-2mm}
	}

	\maketitle
	
	\begin{abstract}
		A dual-scale deployment (DSD) framework is proposed for pinching antenna systems (PASS), under which four implementation protocols are developed. In the proposed framework, coarse-scale deployment moves the pinching antenna (PA) over a wide range along the waveguide, while fine-scale deployment adjusts the PA with high precision within a local region. By jointly optimizing these two scales, the DSD framework fully exploits the flexibility of PA deployment while maintaining low computational complexity. Based on this framework, we establish a practical power-consumption model and derive closed-form expressions for the energy efficiency of PASS. An energy-efficiency maximization problem is then formulated to jointly optimize transmit precoding, PA radiation power, and dual-scale PA deployment. To solve this non-convex and highly coupled problem, a low-complexity penalty-based alternating optimization algorithm is proposed. Simulation results validate the accuracy of the theoretical analysis and the convergence of the proposed algorithm. The proposed DSD framework delivers about $70\%$ higher energy efficiency than the conventional cell-free architecture and nearly a \emph{twofold} improvement over MIMO systems.

		\begin{IEEEkeywords}
			Beamforming, dual-scale antenna deployment, energy efficiency, pinching antenna system, protocol design.
		\end{IEEEkeywords}

	\end{abstract}

	\section{Introduction}
	\IEEEPARstart{T}{he} advent of sixth-generation (6G) wireless systems is anticipated to impose far more stringent requirements on spectral efficiency and full-dimensional coverage\cite{6G}. To meet these demands, emerging technologies such as reconfigurable intelligent surfaces \cite{RIS}, fluid antennas \cite{fluid1}, and movable antennas \cite{movable1} have shown promising performance improvements. These flexible antenna technologies advance transceiver architectures and enable environmental reconfigurability, which allows for controllable signal shaping. The potential benefits of these technologies, however, are limited because they do not fully mitigate the performance degradation caused by large-scale fading. Recently, the pinching antenna system (PASS) \cite{PA_mag, PA_ding, tutorial} has attracted significant attention because of its ability to boost spectral efficiency through near-wire communications. Specifically, in PASS, the dielectric waveguide serves both as the transport medium and the antenna rail, while small dielectric particles are pinched onto the waveguide to form pinching antennas (PAs), which act as programmable leakage points for signal radiation. Compared with prior flexible-antenna approaches, PASS directs part of the signal propagation through low-attenuation waveguides and radiates it closer to users, thereby reshaping large-scale fading and significantly reducing path loss. This additional spatial degree of freedom (DoF) allows the optimization of PA deployment and power radiation, enabling PA beamforming to shape the emitted field. These capabilities enhance communication quality, for example by facilitating unobstructed communication links~\cite{LoS_links} and improving physical-layer security \cite{security1,security2}. Therefore, PASS has drawn tremendous attention from both academia and industry\cite{docomo}, and is expected to play an important role in future wireless communications.

	\subsection{Prior Works}
	Motivated by the aforementioned favorable characteristics of PASS, substantial research has been conducted to optimize the design of PA power radiation and deployment \cite{PA_discrete1,PA_discrete2,PA_general,model_PA,PA_equal1}, aiming to fully exploit the additional DoFs introduced by PA and achieve performance gains in PASS.
	
	More specifically, the authors of \cite{model_PA} introduced both equal-power and proportional-power radiation schemes under a physics-based hardware model, formulating a transmit power minimization problem that jointly optimizes transmit and pinching beamforming. Due to its simplicity, the equal-power radiation approach was adopted in \cite{PA_equal1}, where it was applied to multicast scenarios, with pinching beamforming designed to maximize the multicast rate. In contrast, the general-power scheme offers more flexibility in power allocation across PAs with the potential for higher performance, although its hardware implementation is more complex. For instance, a general-power radiation scheme was proposed in~\cite{PA_general}, yielding significant transmit power savings compared to the equal-power approach. In terms of PA deployment optimization, a practical strategy involves pre-deploying a large number of PAs and discretely activating a subset to serve communication users. In this regard, the authors of \cite{PA_discrete1} formulated a transmit power minimization problem that optimizes transmit beamforming, pinching beamforming, and the number of active PAs, while ensuring the rate constraint of individual users is met. Additionally, the authors of \cite{PA_discrete2} proposed three practical PA placement schemes, i.e., edge-deployment, center-deployment, and diagonal-deployment. Furthermore, continuous PA activation methods have been investigated in \cite{PA_general,model_PA,PA_equal1} to maximize system performance by optimizing PA deployment within the waveguide region, subject to the constraint that the distance between any two PAs exceeds a minimum threshold.

	\subsection{Motivations and Contributions}
	Although extensive studies have explored the substantial performance gains from continuous PA deployment, its practical adoption still faces two key concerns. First, under the existing hardware platforms, the practical implementation of continuous PA deployment cannot achieve the ideal combination of both wide-range and high-precision deployment. This limitation arises because the motor-driven device inherently involves a trade-off between travel range and spatial precision within a limited response time. Second, the power consumption associated with continuous PA deployment can be substantial, which may degrade the overall energy efficiency.

	To effectively address these concerns, we propose a dual-scale deployment (DSD) framework that employs a hierarchical design for PA deployment. In the proposed framework, each PA is physically housed in a mobile base that slides on the waveguide, while the PA itself is tunable within the base. This architecture naturally transforms the PA deployment into dual-scale operations, ensuring both wide-range repositioning and high-precision refinement. A similar framework is reflected in the two-stage algorithms adopted in \cite{two_stage1,two_stage2,two_stage3}, but these works rely on the decoupled optimization of these two scales. For instance, \cite{two_stage1} minimized the path loss by sequentially moving the PA to the nearest position to the user before phase tuning. However, such decoupled optimization fails to effectively generalize to multi-user interference scenarios, since it overlooks the spatial coupling between the two scales essential for mitigating inter-user interference. To bridge this gap, we design an energy-efficiency maximization algorithm to jointly optimize the transmit precoding, PA power radiation, and dual-scale PA deployment. The main contributions of this paper are summarized as follows:
	\begin{itemize}
		\item We propose the DSD framework for PA deployment in PASS. In the proposed DSD framework, the mobile base executes coarse-scale sliding, while the PA performs fine-scale tuning. We provide four implementation protocols within the DSD framework, namely: ``base sliding then PA tuning (STT)'', ``base selection then PA activation (STA)'', ``base selection and PA tuning (SAT)'', and ``base selection and PA activation (SAA)'', along with their respective advantages and limitations.

		\item We propose a realistic PASS power consumption model. Based on the proposed model, we derive closed-form theoretical expressions for energy efficiency in the DSD framework. Then, we formulate an energy efficiency maximization problem to jointly optimize the transmit precoding, PA radiation power, and dual-scale PA deployment under four implementation protocols.

		\item We propose a novel low-complexity algorithm to efficiently address the energy-efficiency maximization problem. In particular, we decompose the dual-scale PA deployment optimization in the DSD framework into two subproblems, i.e., coarse-scale base placement and fine-scale PA deployment. The proposed algorithm also incorporates Lagrangian transformations and block coordinate manifold optimization to tackle the transmit precoding and PA power radiation problems.

		\item We provide numerical results to validate the accuracy of derived theoretical results and the convergence of the proposed algorithm. Numerical results also demonstrate that: 1) the proposed DSD framework is highly effective for PASS, delivering about $70\%$ higher energy efficiency than the conventional cell-free architecture and nearly a twofold improvement relative to MIMO systems; 2) the impact of DSD resolutions on energy efficiency is more pronounced in multi-PA systems; 3) the STT protocol provides over $80\%$ improvement in energy efficiency due to the sliding-tuning gain, while the SAT protocol delivers the highest energy efficiency among the four protocols.

	\end{itemize} 
	
	\subsection{Organization and Notations}
	The remainder of this paper is structured as follows. Section~\ref{sec:DSP} provides the signal model for PASS and four implementation protocols in the DSD framework. Section~\ref{sec:model} introduces the system model and formulates the energy efficiency maximization problem for PASS. Section~\ref{sec:EE_maximization} details the proposed penalty-based alternating algorithm for joint optimization of the transmit precoding, PA power radiation, and PA deployment under the four implementation protocols. Numerical results evaluating the performance of different methods under various system configurations are presented in Section~\ref{sec:simulation}. Finally, Section \ref{sec:conclusion} concludes the paper.
	
	\emph{Notations:} Scalars, vectors, and matrices are denoted by lower-case, bold-face lower-case, and bold-face upper-case letters, respectively. The sets of complex and real numbers are represented by $\mathbb{C}$ and $\mathbb{R}$, respectively. The transpose, conjugate transpose, inverse, and trace operations are represented by $(\cdot)^T$, $(\cdot)^H$, $(\cdot)^{-1}$, and $\mathrm{tr}(\cdot)$, respectively. The absolute value and Euclidean norm are indicated by $|\cdot|$ and $\|\cdot\|$, respectively. The symbol $\circ$ denotes element-wise multiplication. The expectation operator is denoted by $\mathbb{E}[\cdot]$. The real part of a complex number is denoted by $\Re \{\cdot\}$. $\mathbf{I}_N$ denotes an identity matrix of size $N \times N$, and $\mathbf{1}_N  = [1,1,\cdots,1]^T \in \mathbb{R}^{N \times 1}$. $[\mathbf{A}]_{m,n}$ represents the element in the $m$-th row and $n$-th column of $\mathbf{A}$. The operator $\mathrm{blkdiag}\{\mathbf{x}_1,\mathbf{x}_2,\cdots,\mathbf{x}_M  \}$ denotes a block diagonal matrix with diagonal blocks $\mathbf{x}_1,\mathbf{x}_2,\cdots,\mathbf{x}_M$.
	
	\section{Dual-Scale Deployment Protocols for PASS}\label{sec:DSP}
	In this section, we first present the basic signal model for PASS, based on which four distinct implementation protocols in the DSD framework for PA deployment are introduced.
	
	\subsection{Signal Model for PASS}
	
	\begin{figure}[t]
		\centering
		\includegraphics[width=0.9\linewidth]{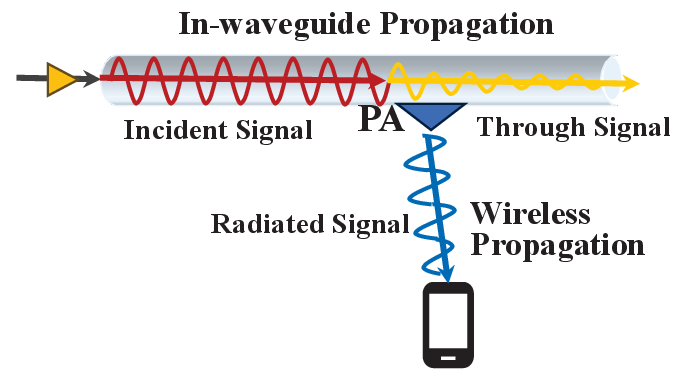}
		\caption{The concept of PASS signal propagation.}
		\label{fig_pa_radiate}
	\end{figure}
	
	Fig.~\ref{fig_pa_radiate} illustrates a simple PASS featuring a waveguide and a PA, where the incident signal into the waveguide is divided into the radiated and through signals at the PA. To characterize this PA feature, let $s_{n-1}^t$ denote the signal transmitted towards the $n$-th PA, where $n \in \{1,2,\cdots,N\}$, $s_0^t$ denotes the original signal fed into the waveguide, and $N$ denotes the total number of PAs. In particular, the signal $s_{n-1}^t$ will travel through the waveguide of distance $L^i_n$ and reach the $n$-th PA, which is modeled as the incident signal $s_n^i$. Due to the finite conductor resistance and dielectric loss of the waveguide, the mathematical relationship between $s_{n-1}^t$ and $s_n^i$ can be modeled as the following energy attenuation expression [Eq.~(8-47),~\citen{EM_field}] as
	\begin{equation}
		s_n^i = \exp\left( -\gamma_g L_n^i \right) \cdot s_{n-1}^t.
	\end{equation}
	Here $\gamma_g$ is the in-waveguide attenuation constant defined as $\gamma_g = \alpha_g + j \beta_g$, where $\alpha_g$ and $\beta_g$ denote the positive-value amplitude and phase constants, respectively. The values of $\alpha_g$ and $\beta_g$ are determined by waveguide parameters and the signal wavelength, which can be obtained via the Eqs.~(8-48) and (8-49) of the work \cite{EM_field}, respectively. In particular, $\beta_g$ can be approximated by $\omega_g \sqrt{\mu_g \epsilon_g}$ \cite{EM_field}, consistent with the in-waveguide signal model \cite{PA_ding} as $\beta_g = \frac{2\pi}{\lambda_g}$, where $\lambda_g=\frac{2\pi}{ \omega_g \sqrt{\mu_g \epsilon_g}}$ is the effective signal wavelength in the waveguide medium. As such, the incident signal at the $n$-th PA is expressed as
	\begin{equation}\label{s1}
		s_n^i = \exp\left( -\left(\alpha_g + j \frac{2\pi}{\lambda_g} \right) L_n^i \right) s_{n-1}^t.
	\end{equation}  
	
	The $n$-th PA attached to the waveguide radiates part of the energy through electromagnetic coupling, while the remaining signal continues to transmit through the waveguide. Let $s_n^r$ and $s_n^t$ represent the radiated and through signals at the $n$-th PA, respectively, and can be modeled \cite{model_PA} as
	\begin{equation}\label{s2}
		s_n^r = \sqrt{\delta_n} s_n^i,
	\end{equation}         
	where $\delta_n \in [0,1]$. According to the law of energy conservation, the sum of the power of radiated and through signals has to be equal to the incident signal power, i.e., $|s_n^r|^2 + |s_n^t|^2 = |s_n^i|^2$, which yields
	\begin{equation}\label{s3}
		s_n^t = \sqrt{1-\delta_n} s_n^i.
	\end{equation}
	
	Then, the through signal propagates towards the $(n+1)$-th PA along the waveguide and serves as the next incident signal $s_{n+1}^i$. On the other hand, $s_n^r$ is transmitted via wireless propagation to serve communication users. Let $h_{nk}$ denote the wireless channel between the $n$-th PA and the $k$-th single-antenna user (or $\mathbf{h}_{nk}$ for the multi-antenna user). The received signal at the $k$-th user from the $n$-th PA radiation is given by
	\begin{equation}\label{y1}
		y_{nk} = h_{nk} s_n^r + n_k,
	\end{equation}
	where $n_k$ is the additive white Gaussian noise $\mathcal{CN}(0,N_0)$ and $N_0$ is the noise power. As a simple example, the wireless channels in PASS can be modeled by a free-space line-of-sight (LoS) channel as~\cite{PA_ding,model_PA}
	\begin{equation}\label{h1}
		h_{nk} = \sqrt{\frac{c^2}{16 \pi^2 f_c^2}} \frac{e^{-j \frac{2\pi}{\lambda} r_{nk}}}{r_{nk}},
	\end{equation}
	where $c$ is the light speed, $f_c$ and $\lambda$ are the carrier frequency and wavelength, respectively, and $r_{nk}$ is the distance between the $n$-th PA and $k$-th user. Substituting Eqs.~\eqref{s1}-\eqref{s3}, and \eqref{h1} into~\eqref{y1}, the received signal at the $k$-th user from $N$ PAs is further expressed as
	\begin{align}\label{r1}
		y_{k} =& \sum_{n=1}^N \sqrt{\frac{c^2}{16 \pi^2 f_c^2}} \frac{\sqrt{\xi_n}}{r_{nk}} \nonumber \\
		& \times \exp\left( -\alpha_g d_n -j \left(\frac{2\pi}{\lambda} r_{nk} +  \frac{2\pi}{\lambda_g} d_n \right)  \right) s_0^t  + n_k.
	\end{align}
	where $\xi_n = \delta_n \prod_{j=1}^{n-1}(1 - \delta_j)$ and $d_n = \sum_{j=1}^n L_j^i$ denote the overall radiation power coefficient and in-wavelength propagation length for the $n$-th PA. Let $P^\mathrm{d}_k$ denote the received desired-signal power of the $k$-th user, where 
	\begin{equation}\label{Pk}
		P^\mathrm{d}_k \!=\! \left| \sum_{n=1}^N\!\! \sqrt{\frac{c^2}{16 \pi^2 f_c^2}}\! \frac{\sqrt{\xi_n}}{r_{nk}}\! \exp\!\! \left( \!-\alpha_g d_n \!-\! j\! \left(\! \frac{2\pi}{\lambda} r_{nk}\! + \! \frac{2\pi}{\lambda_g} d_n\! \right) \! \right)\! \right|^2\!\!.
	\end{equation}
	
	This expression shows that $P^\mathrm{d}_k$ is jointly determined by the PA power radiation coefficients $\{\xi_n\}_{n=1}^N$ and the PA deployment through $\{r_{nk}, \ d_n \}_{n=1}^N$. Thus, maximizing the received desired-signal power requires joint optimization of PA power radiation and PA deployment. Strategies for allocating PA radiation power, including the equal- and proportional-power models, have been introduced and analyzed in \cite{model_PA}. In contrast, practical protocols for adjusting PA position remain underexplored. To bridge this gap, in the following part, we develop four implementation protocols in the DSD framework for PA deployment in PASS.

	\vspace{-3mm}
	\subsection{Four Implementation Protocols in the DSD Framework}
	As shown in~\eqref{Pk}, the deployment of the $n$-th PA simultaneously affects the amplitude and phase of $P_k^{\mathrm{d}}$, through $d_n$ and $r_{nk}$. Thus, it is very difficult to obtain the optimal PA deployment that maximizes $P_k^{\mathrm{d}}$, as the underlying optimization is highly non-convex and exhibits strong coupling among the $N$ PAs. To solve these challenges, we divide the effect of PA deployment on $P_k^{\mathrm{d}}$ into amplitude and phase components, respectively, as 
	\begin{align}
		a_{nk}(\mathbf{p}_n) & = \left( r_{nk}(\mathbf{p}_n) \right)^{-1} \exp\left( -\alpha_g d_n(\mathbf{p}_n) \right), \\
		\theta_{nk}(\mathbf{p}_n) & =  \frac{2\pi}{\lambda} r_{nk}(\mathbf{p}_n) +  \frac{2\pi}{\lambda_g} d_n(\mathbf{p}_n),  
	\end{align}
	where $\mathbf{p}_n$ denotes the position of the $n$-th PA. Then, the desired-signal power problem with respect to (w.r.t.) PA deployment can be rewritten as
	\begin{equation}\label{sec1_op}
		\max_{\mathbf{p}_n \in \bm{\mathcal{G}}_n} \quad \left| \sum_{n=1}^N \sqrt{\frac{c^2 \xi_n}{16 \pi^2 f_c^2}} a_{nk}(\mathbf{p}_n) \exp\left( -j \theta_{nk}(\mathbf{p}_n) \right) \right|^2,
	\end{equation}  
	where $\bm{\mathcal{G}}_n$ denotes the deployment range of the $n$-th PA along the waveguide.
	
	It can be observed that the challenge of solving  problem~\eqref{sec1_op} lies in the coupling between $a_{nk}$ and $\theta_{nk}$, $\forall n$. To solve this, we propose a dual-scale optimization method. The insight behind this method is that a small variation of the PA position $\Delta \mathbf{p}_n$ induces a negligible change in the amplitude term $a_{nk}(\mathbf{p}_n + \Delta \mathbf{p}_n)$, but enables $\theta_{nk}(\mathbf{p}_n + \Delta \mathbf{p}_n)$ to sweep the full range of $[-\pi,\pi)$.\footnote{Under the parameter setting in Table~\ref{simu}, consider the channel between a PA at $\mathbf{p}_n=[98.6,75]^T$ on the third waveguide and user~2. A PA displacement of $7.65$ mm changes $\theta_{n2}$ by approximately $2\pi$. Meanwhile, the distance-dependent factor $r_{n2}^{-1}$ in $a_{n2}$ changes from $1/10.4$ to $1/\sqrt{10.4^2+0.00765^2}$, whose relative variation is below $3\times 10^{-7}$.}Hence, $\theta_{nk}$ can be tuned by small PA movements with negligible amplitude perturbation, which validates the decoupling approximation and motivates the DSD framework. In this way, we first solve the decoupled problem as
	\begin{equation}\label{sec1_op2}
		\max_{a_{nk}, \theta_{nk}} \quad \left| \sum_{n=1}^N \sqrt{\frac{c^2 \xi_n}{16 \pi^2 f_c^2}} a_{nk} \exp\left( -j \theta_{nk} \right) \right|^2,
	\end{equation}   
	where the range of variable $a_{nk}$ is determined by the position of $k$-th user and $\bm{\mathcal{G}}_n$.

	Based on the optimal $a_{nk}^{\mathrm{opt}}$ and $\theta_{nk}^{\mathrm{opt}}$ derived from the problem~\eqref{sec1_op2}, the proposed DSD strategy carries out coarse- and refinement-deployment of the $n$-th PA in a dual-scale manner. Particularly, for the coarse-scale deployment, the $n$-th PA positioning is adjusted to $\mathbf{p}_n^C$ to achieve the desired amplitude by solving
	\begin{equation}\label{pC}
		\mathbf{p}_n^C = \argmin_{\mathbf{p}_n \in \bm{\mathcal{G}}_n} \left| a_{nk}^{\mathrm{opt}} - \left( r_{nk}(\mathbf{p}_n) \right)^{-1} \exp\left( -\alpha_g d_n(\mathbf{p}_n) \right) \right|^2.
	\end{equation}  
	
	Then, for the fine-scale deployment, the $n$-th PA will be further refined with $\Delta \mathbf{p}_n^F$ to achieve the desired phase by solving
	\begin{align}\label{pF}
		\Delta \mathbf{p}_n^F = & \argmin_{\Delta \mathbf{p}_n \in \Delta \bm{\mathcal{G}}_n} 
		\Bigg| \theta_{nk}^{\mathrm{opt}} - \bigg( \frac{2\pi}{\lambda} r_{nk}(\mathbf{p}_n^C \nonumber \\
		& + \Delta \mathbf{p}_n) +  \frac{2\pi}{\lambda_g} d_n(\mathbf{p}_n^C + \Delta \mathbf{p}_n)  \bigg)  \Bigg|^2,
	\end{align}
	where $ \Delta \bm{\mathcal{G}}_n$ denotes the deployment range of the $n$-th PA within the base. After the proposed DSD operations, the final optimized position of the $n$-th PA is $\mathbf{p}_n^F = \mathbf{p}_n^C + \Delta \mathbf{p}_n^F$. 
	
	\begin{remark}
		In the proposed DSD framework, the $n$-th PA is first transferred along the waveguide to a wide-range waypoint $\mathbf{p}_n^C$ to mitigate large-scale path-loss, typically over several to tens of meters. It then performs a high-precision refinement to the final location $\mathbf{p}_n^F$ to ensure small-scale fading-phase alignment, typically within several micrometers to a few millimeters. The proposed DSD framework reformulates PA deployment into two decoupled subproblems, which substantially reduces the computational complexity of the PASS designs. It can also be smoothly applied to a motorized actuator system to ensure both wide-range repositioning and high-precision refinement.
	\end{remark}
	
	In practice, the proposed DSD framework can be realized with a hardware platform from Physik Instrument that integrates motorized and piezoelectric actuators~\cite{PI}. The motorized module enables wide-range movement to the coarse position, while the piezoelectric actuator provides precise fine tuning to the final position. Accordingly, the $n$-th PA requires an operation time $t_n = \frac{\|\mathbf{p}_n^C\|}{v_n^{MO}} + \frac{\|\Delta \mathbf{p}_n^F \|}{v_n^{PI}}$ during the position-adjustment process, where $v_n^{MO}$ and $v_n^{PI}$ denote the speeds of the motorized and piezoelectric actuators, respectively. With representative values $v_n^{MO} = 10$ m$/$s and $v_n^{PI} = 10^{-2}$ m$/$s, this operation time far exceeds the $10$ ms NR frame duration~\cite{NR_frame}. Hence, practical PA deployment has a latency on the order of seconds and therefore cannot support real-time adaptation to communication channels.

\begin{figure}[t]
		\centering
		\includegraphics[width=0.9\linewidth]{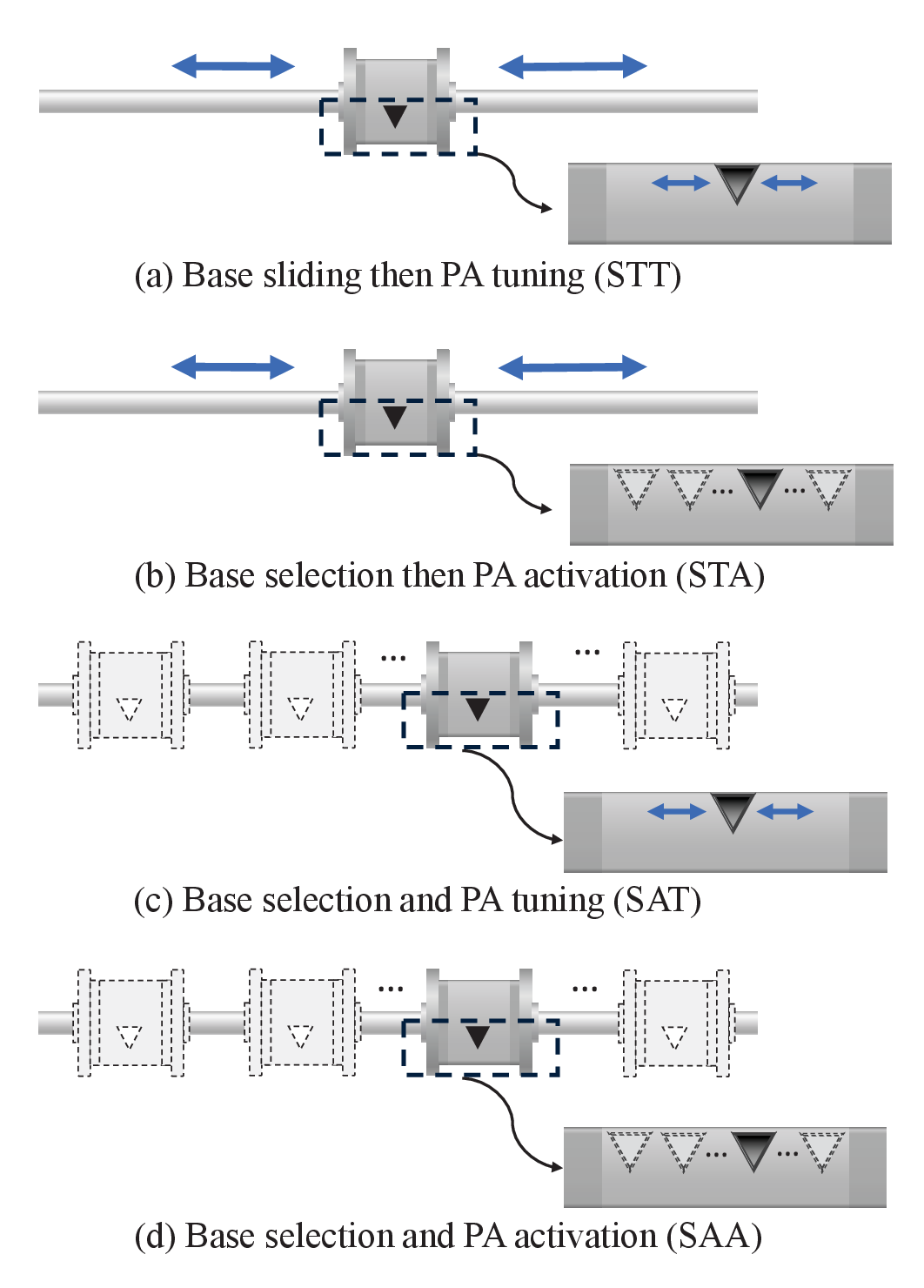}
		\caption{Schematic diagram of four protocols in the proposed DSD framework.}
		\label{pa_protocol}
	\end{figure}

	Fortunately, one simple and effective solution is to pre-deploy a large number of PAs and then activate a subset to achieve the same effect as adjusting the PA positions. Nevertheless, this also raises concerns about the performance degradation and hardware cost associated with pre-deploying a large number of PAs. For example, implementing centimeter-level positioning accuracy over $100$ m would require $10^4$ pre-deployed PAs. To strike a satisfactory trade-off among the number of pre-deployed PAs, positioning accuracy, and response time, we introduce four implementation protocols within the DSD framework, as shown in Fig.~\ref{pa_protocol}.
	
	\begin{itemize}
		\item \textbf{STT:} Base sliding is carried out by a motorized module over a waveguide-level range for coarse-scale deployment. Fine-tuning PA inside the base is then performed by a piezoelectric module.
		
		\item \textbf{STA:} Base sliding is carried out by a motorized module over a waveguide-level range for coarse-scale deployment. $N^F$ PAs are pre-deployed inside the base, where one PA is activated each time slot.
		
		\item \textbf{SAT:} $N^C$ bases are pre-deployed along the waveguide, each equipped with a tunable PA. In each time slot, one base is selected, and the PA within that base is finely tuned by a piezoelectric module.
		
		\item \textbf{SAA:} $N^C$ bases are pre-deployed along the waveguide, each containing $N^F$ PAs. In each time slot, one base is selected, and one PA within that base is activated.
	\end{itemize}
	
	\setlength{\arrayrulewidth}{0.2mm} 
	\renewcommand{\arraystretch}{1.2}

	\begin{table*}[htbp]
		\centering
		\begin{tabular}{|>{\centering\arraybackslash}m{4cm}|>{\centering\arraybackslash}m{2.3cm}|>{\centering\arraybackslash}m{2cm}|>{\centering\arraybackslash}m{3cm}|>{\centering\arraybackslash}m{3cm}|}
			\hline
			\rowcolor[HTML]{E6E6E6}
			\textbf{Protocol} & \textbf{Response Time
			} & \textbf{Pre-deployed PA Number} & \textbf{Power Consumption} & \textbf{Spectral Efficiency} \\
			\hline 	
			STT  & Long & Small & Very High & High \\ \hline	
			STA  &   Moderate &   Moderate & High &Moderate \\ \hline	
			SAT  &  Moderate &  Moderate &  Low &Moderate \\ \hline	
			SAA  & Nearly zero & Large & Very Low &Low \\ \hline	
		\end{tabular}
		\caption{Summary of performance metrics for four PA operating protocols.}
		\label{tab_PA_protocol}
		\vspace{-4mm}
	\end{table*}

	In Table~\ref{tab_PA_protocol}, we summarize the response time, pre-deployed PA number, power consumption, and spectral efficiency of these four implementation protocols~(a)-(d). For protocol selection, STT is suitable for slowly varying scenarios that prioritize spectral efficiency, since it supports both coarse sliding and fine tuning. SAT provides a balanced design, since pre-deployed bases avoid repeated motorized sliding while preserving fine adjustment. SAA is suitable for delay-critical scenarios, at the cost of a large pre-deployment budget and lower spectral efficiency. STA mainly serves as an intermediate option when coarse repositioning is required but fine tuning is not adopted. Across the four implementation protocols, the PA deployment algorithm designs only differ in the search grid sets in Eqs. \eqref{pC} and \eqref{pF}. We next detail the protocol-specific grid constructions. For simplified mathematical modeling, we assume that the waveguide is deployed along the $x$-axis. This means that the PA deployment designs only require consideration of its $x$-coordinate. Let $x^{\min}$ and $x^{\max}$ denote the minimum and maximum coordinates of the base deployment range along the waveguide, respectively. We use $\delta^C$ and $\delta^F$ to indicate the position adjustment step sizes, or resolutions, of the sliding base and tuning PA, respectively. Besides, let $L^F$ denote the length of PA's deployment range within each base for the refinement. As for the selection or activation representations, it is assumed that $N \cdot N^F$ bases and $N \cdot N^C$ PAs are pre-deployed, from which $N$ bases or PAs are selected/activated in each time slot. Four practical implementations of the proposed DSD search grid set are provided as follows:
	
	\begin{itemize}
		\item \textbf{STT:}
		\begin{align}\label{dsp1}
			& \mathbf{x}^{CG} = \left\{ x^C \Bigg| \begin{array}{l}
				x^C = x^{\min} +g^C \delta^C,\\
				g^C=0,1,\cdots,\frac{x^{\max}-x^{\min}}{\delta^C}
			\end{array}  \right\}, \\
			& \mathbf{x}^{FG} = \left\{ x^F \Bigg|  \begin{array}{l}
				x^F = -\frac{L^F}{2} + g^F \delta^F,\\
				g^F = 0,1,\cdots,\frac{L^F}{\delta^F}
			\end{array} \right\}.
		\end{align}
		
		\item \textbf{STA:}
		\begin{align}
			&\mathbf{x}^{CG} = \left\{ x^C \Bigg| \begin{array}{l}
				x^C = x^{\min} +g^C \delta^C,\\
				g^C=0,1,\cdots,\frac{x^{\max}-x^{\min}}{\delta^C}
			\end{array}  \right\}, \\
			& \mathbf{x}^{FG} =\left\{x^F \Bigg| \begin{array}{l}
				x^F = -\frac{L^F}{2} +g^F \frac{L^F}{N N^F},\\
				g^F =0,1, \cdots,N N^F - 1  
			\end{array}  \right\}.
		\end{align}
		
		\item \textbf{SAT:} 
		\begin{align}
			& \mathbf{x}^{CG}= \left\{ x^C \Bigg|  \begin{array}{l}
				x^C =x^{\min} + g^C \frac{x^{\max}- x^{\min}}{N N^C},\\
				g^C =0,1, \cdots ,N N^C-1 
			\end{array} \right\}, \\
			& \mathbf{x}^{FG} = \left\{ x^F \Bigg|  \begin{array}{l}
				x^F = -\frac{L^F}{2} + g^F \delta^F,\\
				g^F = 0,1,\cdots,\frac{L^F}{\delta^F}
			\end{array} \right\}.
		\end{align}
		\item \textbf{SAA:}
		\begin{align}\label{dsp2}
			& \mathbf{x}^{CG}= \left\{ x^C \Bigg|  \begin{array}{l}
				x^C =x^{\min} + g^C \frac{x^{\max}- x^{\min}}{N N^C},\\
				g^C =0,1, \cdots ,N N^C-1 
			\end{array} \right\}, \\
			&  \mathbf{x}^{FG} =\left\{x^F \Bigg| \begin{array}{l}
				x^F = -\frac{L^F}{2} +g^F \frac{L^F}{N N^F},\\
				g^F =0,1, \cdots,N N^F - 1  
			\end{array}  \right\}.
		\end{align}
	\end{itemize}
	Here, the analytical expressions of $\mathbf{x}^{CG}$ and $\mathbf{x}^{FG}$ correspond to the $x$-axis of $\bm{\mathcal{G}}_n$ and $\Delta \bm{\mathcal{G}}_n$ in Eqs.~\eqref{pC} and \eqref{pF}, respectively, and can be easily extended to two or three dimensions. It is crucial to identify the optimal protocols~(a)-(d) to achieve the best performance of PASS. In the subsequent sections, we will jointly optimize the transmit precoding, PA radiation power, and PA deployment under the four implementation protocols to maximize the energy efficiency of PASS.

	\section{System Model and Problem Formulation}\label{sec:model}
	Based on the basic signal model and the proposed DSD framework for PASS-enabled wireless communications, this section presents a general system model in which $M$ waveguides are fed by $M$ dedicated RF chains, each consisting of $N$ PAs, to cooperatively serve $K$ users. Let $\mathbf{q}_k=[x_k^U, y_k^U]^T$ and $\mathbf{p}_{mn}$ denote the position coordinates of the $k$-th user and the $n$-th PA on the $m$-th waveguide, respectively. The following subsection provides a detailed analysis of the transmit signal with RF-chain precoding, the received signal at the user side, and the system's energy efficiency.

	\subsection{Signal Model}
	Let $s_k$ denote the information symbol for the $k$-th user with $\mathbb{E}\{|s_k|^2\}=1$. After the $m$-th RF chain precodes $s_k$ with weight $w_{mk}$, $\forall k$, for the $K$ users, the resulting signal fed into the $m$-th waveguide is given by
	\begin{equation}
		x_m = \sum_{k=1}^K w_{mk} s_k,
	\end{equation}
	where the total transmit power of $M$ RF chains is constrained by $P_{\max}$, i.e., $\mathbb{E}\left\{\sum_{m=1}^M |x_m|^2\right\}  \le P_{\max}$.

	The received signal of the $k$-th user comprises the radiated signal from the $N$ PAs along each of the $M$ waveguides, i.e.,
	\begin{equation}\label{yk}
		y_k = \sum_{m=1}^M \sum_{n=1}^N \sqrt{\xi_{mn}} g_{mn} h_{mnk} x_m + n_k,
	\end{equation}     
	where $g_{mn}$ and $h_{mnk}$ denote the in-waveguide channel of the $n$-th PA on the $m$-th waveguide, and wireless channel from the $n$-th PA on the $m$-th waveguide to the $k$-th user, respectively. $\xi_{mn}$ is the radiation power coefficient of the $n$-th PA on the $m$-th waveguide, satisfying $\sum_{n=1}^N \xi_{mn} = 1$, $\forall m$, and $n_k \sim \mathcal{CN}(0,N_0)$ is the additive white Gaussian noise. More particularly, the in-waveguide channel is given by (c.f.~\eqref{s1})
	\begin{equation}\label{gmn}
		g_{mn} = \exp\left( -\left(\alpha_g + j \frac{2\pi}{\lambda_g} \right) d_{mn} \right),
	\end{equation}
	where $d_{mn}$ is in-waveguide propagation distance for the $n$-th PA on the $m$-th waveguide. The wireless channel model combines both the LoS and NLoS links, i.e.,
	\begin{equation}\label{hmnk}
		h_{mnk} \!=\! \sqrt{C_0 r_{mnk}^{-\beta_u}} \!\left(\! \sqrt{\frac{K_r}{K_r+1}} e^{-j \frac{2\pi}{\lambda} r_{mnk}} \!+\! \sqrt{\frac{1}{K_r+1} } \tilde{h}_{mnk} \!\right),
	\end{equation}
	where $C_0$ and $\beta_u$ denote the path-loss coefficient at the reference distance $1$ m and the path-loss exponential coefficient, respectively, $r_{mnk}$ is the distance from the $n$-th PA on the $m$-th waveguide to the $k$-th user, $K_r \ge 0$ is the Rician factor to measure the power ratio between the LoS and NLoS links, and $\tilde{h}_{mnk} \sim \mathcal{CN}(0,1)$ represents NLoS links. It can be observed that the free-space LoS channel in~\eqref{h1} is a special case of our considered channel in~\eqref{hmnk}, and becomes equivalent when $K_r \to +\infty$, $C_0 = \frac{c^2}{16 \pi^2 f_c^2}$ and $\beta_u=2$.

	\subsection{Energy Efficiency}
	In practical communication systems, it is typically assumed that the BS has perfect channel state information (CSI), while users only possess statistical CSI. In this case, the effective SINR of the $k$-th user is given by~\cite{cellfree_model}:	
	\begin{equation}\label{gammak1}
		\gamma_k = \frac{\left|\text{DS}_k\right|^2}{\mathbb{E}\left\{ |\text{CU}_k |^2 \right\} + \sum_{i \neq k}^K \mathbb{E} \left\{ | \text{UI}_{ki} |^2 \right\} + N_0  },
	\end{equation} 
	where $\text{DS}_k$, $\text{CU}_k$, and $\text{UI}_{k i}$ denote the $k$-th user's desired signal, effective interference from the CSI uncertainty, and the other $i$-th user interference signals, respectively, given by
	\begin{align}
		& \text{DS}_k = \mathbb{E}\left\{ \sum_{m=1}^M \sum_{n=1}^N \sqrt{\xi_{mn}} g_{mn} h_{mnk} w_{mk} \right\}, \\
		& \text{CU}_k = \sum_{m=1}^M \sum_{n=1}^N \sqrt{\xi_{mn}} g_{mn} h_{mnk} w_{mk}   -  \text{DS}_k, \\
		& \text{UI}_{k i} = \sum_{m=1}^M \sum_{n=1}^N \sqrt{\xi_{mn}} g_{mn} h_{mnk} w_{mi} .
	\end{align}
	Specifically, $\text{DS}_k$ is the average effective channel gain for decoding $s_k$, $\text{CU}_k$ is the mismatch between the instantaneous desired signal and its statistical mean, which is treated as self-interference, and $\text{UI}_{ki}$ accounts for the multi-user interference induced by the $i$-th user's precoded signal. The SINR expression \eqref{gammak1} can be reformulated into a more compact matrix form as follows:
	\begin{equation} \label{gammak2}
		\gamma_k = \frac{\left| \mathbf{\bar{h}}_k^T \bm{\Xi} \mathbf{G}  \mathbf{w}_k \right|^2}{\mathbb{E}\left\{ | \mathbf{\tilde{h}}_k^T \bm{\Xi} \mathbf{G}  \mathbf{w}_k  |^2 \right\} + \sum_{i \neq k}^K \mathbb{E} \left\{ | \mathbf{h}_k^T \bm{\Xi} \mathbf{G}  \mathbf{w}_i  |^2 \right\} + N_0  },
	\end{equation} 
	where $\mathbf{\bar{h}}_k \in \mathbb{C}^{MN \times 1}$, $\mathbf{\tilde{h}}_k \in \mathbb{C}^{MN \times 1}$, and $\mathbf{h}_k \in \mathbb{C}^{MN \times 1}$ are the deterministic, stochastic and overall channels of wireless propagation from $MN$ PAs to the $k$-th user, respectively, defined by
	\begin{align}
		\left[ \mathbf{\bar{h}}_k \right]_{(m-1)N+n} & = \sqrt{C_0 r_{mnk}^{-\beta_u}} \sqrt{\frac{K_r}{K_r+1}} e^{-j \frac{2\pi}{\lambda} r_{mnk}}, \\
		\left[ \mathbf{\tilde{h}}_k \right]_{(m-1)N+n} & = \sqrt{\frac{C_0}{K_r+1}} r_{mnk}^{-\frac{\beta_u}{2}} \tilde{h}_{mnk}, \\
		\left[ \mathbf{h}_k \right]_{(m-1)N+n} &= h_{mnk}.
	\end{align}
	Furthermore, $\mathbf{G} = \mathrm{blkdiag}\left\{\mathbf{g}_1,\mathbf{g}_2,\cdots,\mathbf{g}_M  \right\} \in \mathbb{C}^{MN \times M}$ and $\bm{\Xi} = \mathrm{diag}\{\sqrt{\xi_{11}},\cdots,\sqrt{\xi_{1N}},\sqrt{\xi}_{21},\cdots,\sqrt{\xi_{2N}},\cdots,\sqrt{\xi_{MN}} \} \in \mathbb{R}^{MN \times MN}$ denote the overall in-waveguide channel and PA power radiation matrices, respectively.
	The expectations in \eqref{gammak2} can be calculated by
	\begin{align}
		\mathbb{E}\left\{ \left| \mathbf{\tilde{h}}_k^T \bm{\Xi} \mathbf{G} \mathbf{w}_k  \right|^2 \right\} & = \frac{C_0}{K_r+1} \mathbf{w}_k^T  \mathbf{G}^T \bm{\Xi}^T \mathbf{R}_k \bm{\Xi} \mathbf{G}^*  \mathbf{w}_k^*, \\
		\mathbb{E} \left\{ \left| \mathbf{h}_k^T \bm{\Xi} \mathbf{G}\mathbf{w}_i  \right|^2 \right\} & = \mathbf{w}_i^T  \mathbf{G}^T \bm{\Xi}^T \mathbf{\bar{h}}_k \mathbf{\bar{h}}_k^H \bm{\Xi} \mathbf{G}^* \mathbf{w}_i^* \nonumber \\
		& + \frac{C_0}{K_r+1} \mathbf{w}_i^T  \mathbf{G}^T \bm{\Xi}^T \mathbf{R}_k \bm{\Xi} \mathbf{G}^*  \mathbf{w}_i^*,
	\end{align}
	where $\mathbf{R}_k = \mathrm{diag}\Big\{ r_{11k}^{-\beta_u}, r_{12k}^{-\beta_u}, \cdots,r_{1Nk}^{-\beta_u},\cdots,r_{m1k}^{-\beta_u},r_{m2k}^{-\beta_u},\\ \cdots,r_{mNk}^{-\beta_u},\cdots,r_{M1k}^{-\beta_u},r_{M2k}^{-\beta_u},\cdots,r_{MNk}^{-\beta_u} \Big\} \in \mathbb{R}^{MN \times MN}$. Then, the achievable rate of the $k$-th user is given by \eqref{Rk}.
	\begin{figure*}[ht]
		\begin{equation}\label{Rk}
			R_k = \log_2 \left(1+ \frac{\left| \mathbf{\bar{h}}_k^T \bm{\Xi} \mathbf{G}  \mathbf{w}_k \right|^2}{\frac{C_0}{K_r+1}  \mathbf{w}_k^T \mathbf{G}^T \bm{\Xi}^T \mathbf{R}_k \bm{\Xi} \mathbf{G}^*  \mathbf{w}_k^*  + \sum_{i \neq k}^K  \left( \left| \mathbf{\bar{h}}_k^T \bm{\Xi} \mathbf{G}  \mathbf{w}_i \right|^2 + \frac{C_0}{K_r+1}  \mathbf{w}_i^T  \mathbf{G}^T \bm{\Xi}^T \mathbf{R}_k \bm{\Xi} \mathbf{G}^*  \mathbf{w}_i^* \right) + N_0  } \right).
		\end{equation}
		\hrulefill
	\end{figure*}
	
	The power consumption of PASS is primarily from $M$ RF-chains and $M N$ activated PAs, which is expressed as
	\begin{equation}\label{Pall}
		P_{\text{all}} = \sum_{k=1}^K \frac{\|\mathbf{w}_k\|^2}{\nu} + P_{\text{BS}}^{\text{sta}} + MN \cdot P_{\text{PA}},
	\end{equation} 
	where the unique part of the proposed power model lies in $P_{\text{PA}}$, which captures the protocol-dependent PA-side hardware power in PASS, given by
	\begin{equation}
		P_{\mathrm{PA}} =
		\begin{cases}
			P_{\mathrm{PA}}^{\mathrm{act}} + P_{\mathrm{PA}}^{\mathrm{mot}} + P_{\mathrm{PA}}^{\mathrm{pie}}, & \text{for STT},\\
			P_{\mathrm{PA}}^{\mathrm{act}} + P_{\mathrm{PA}}^{\mathrm{mot}},                                  & \text{for STA},\\
			P_{\mathrm{PA}}^{\mathrm{act}} + P_{\mathrm{PA}}^{\mathrm{pie}},                                   & \text{for SAT},\\
			P_{\mathrm{PA}}^{\mathrm{act}},                                    & \text{for SAA}.
		\end{cases}
	\end{equation}
	where $P_{\text{PA}}^{\text{act}}$, $P_{PA}^{\text{mot}}$, and $P_{PA}^{\text{pie}}$ denote the power consumption of PA activation, motorized coarse-scale sliding, and piezoelectric fine-scale tuning, respectively. Based on the derived achievable rate and power consumption model, the energy efficiency of PASS is expressed as
	\begin{equation}\label{eq_EE}
		\eta_{EE} = \frac{B \cdot \sum_{k=1}^K R_k}{P_{\text{all}}},
	\end{equation}
	where $B$ is transmission bandwidth, and $R_k$ and $P_{\text{all}}$ are given by Eqs.~\eqref{Rk} and \eqref{Pall}, respectively.

	\subsection{Problem Formulation}
	Without loss of generality, it can be assumed that the $M$ waveguides are deployed along the $x$-axis. Under this circumstance, the coordinate of the $n$-th PA on the $m$-th waveguide is given by $\mathbf{p}_{mn} = [x_{mn},\ \bar{Y}_{m}]$, where $\bar{Y}_m$ are known a priori, $\forall m$. The distance in channels~\eqref{gmn} and \eqref{hmnk} can be expressed as $d_{mn} = x_{mn}$ and $r_{mnk} = \sqrt{(x_k^U - x_{mn})^2+(y_k^U - \bar{Y}_m)^2}$. The $x$-axis coordinate matrix $\mathbf{X}$ simultaneously affects $\mathbf{G}$ and $\mathbf{h}_k$ for $\forall k$, thereby influencing the energy efficiency of PASS, where $\mathbf{X} = [\mathbf{x}_1 ,\mathbf{x}_2 ,\cdots,\mathbf{x}_M] \in \mathbb{R}^{N \times M}$ and $\mathbf{x}_m = [x_{m1}, x_{m2}, \cdots, x_{mN}]^T$. Under this setup, the EE maximization problem in the DSD framework for PASS can be formulated as
	\begin{subequations}\label{SE1}
		\begin{align}
			\max_{\mathbf{W},\bm{\Xi},\mathbf{X}} \quad & \eta_{EE} \label{P1_EE} \\
			\mathrm{s.t.} \quad &\sum_{k=1}^K \|\mathbf{w}_{k}\|^2 \le P_{\max}, \label{P1_1} \\
			&  \xi_{mn} \ge 0, \ \sum_{n=1}^N \xi_{mn} = 1, \forall m,n \label{P1_2}, \\
			&  x_{mn} \in [x_m^{\min}, x_m^{\max}], \ |x_{mn} \! - \! x_{mn'}| \! \ge \! \Delta x, \ \forall m, n, n', \label{P1_3}
		\end{align}
	\end{subequations}
	where $\mathbf{W} = [\mathbf{w}_1,\mathbf{w}_2,\cdots,\mathbf{w}_K] \in \mathbb{C}^{M \times K}$ is the overall transmit precoding matrix, $x_m^{\min}$ and $x_m^{\max}$ denote the minimum and maximum coordinates of the sliding range of the $m$-th waveguide along the $x$-axis, respectively, and $\Delta x$ is the minimum feasible spacing between any two PAs. The formulated problem presents significant challenges, primarily due to the intricate coupling of optimization variables and the inherent non-convexity of the objective function. First, the optimization variables $\mathbf{W}$, $\bm{\Xi}$, and $\mathbf{X}$ are all mutually coupled in the objective function. Second, the objective function itself is highly non-convex, incorporating fractional and logarithmic terms that further complicate the PASS design. Moreover, the optimization over the $M N$ PA positions $\mathbf{X}$ non-linearly influences both the channel matrices $\{\mathbf{h}_k\}_{k=1}^K$ and $\mathbf{G}$, introducing pervasive non-convexity in both the objective function and the constraints. To overcome these challenges, a penalty-based alternating optimization algorithm is proposed in the following under the DSD framework.

	\section{Energy Efficiency Optimization Design}\label{sec:EE_maximization}
This section will first reformulate the non-convex objective function of problem~\eqref{SE1} into a more tractable form, and then propose a penalty-based alternating optimization algorithm to jointly optimize $\mathbf{W}$, $\bm{\Xi}$, and $\mathbf{X}$ under the DSD framework.

\subsection{Problem Reformulation}
Directly solving problem~\eqref{SE1} is difficult for two main reasons. First, the energy-efficiency objective has a fractional structure, where the sum-rate term is divided by the total power consumption. Second, each user rate contains a logarithmic SINR expression, and the variables $\mathbf{W}$, $\bm{\Xi}$, and $\mathbf{X}$ are still highly coupled in the resulting SINR terms. To address these difficulties, the following lemma introduces auxiliary variables to remove the fractional and logarithmic forms, and equivalently reformulates problem~\eqref{SE1} into a more tractable expression. This reformulated form enables the subsequent alternating-optimization procedure, since it leads to structured subproblems for $\mathbf{W}$, $\bm{\Xi}$, and $\mathbf{X}$.
	
	\begin{lemma}\label{lemma1}
		The energy-efficiency maximization problem can be equivalently transformed into
		\begin{equation}\label{error1}
			\begin{aligned}
				& \min_{\mathbf{W},\bm{\Xi},\mathbf{X}} \ B \cdot \sum_{k=1}^K \kappa_k \epsilon_k + q \cdot P_{all}  \\
				& \mathrm{s.t.} \ \eqref{P1_1}, \ \eqref{P1_2}, \ \eqref{P1_3},
			\end{aligned}
		\end{equation}
		where the optimization variables are				
		\begin{align}\label{epsilon}
			\epsilon_k =& |t_k|^2 \! \sum_{i=1}^K \!\! \left(\!  \left| \mathbf{\bar{h}}_k^T \bm{\Xi} \mathbf{G} \mathbf{w}_i \right|^2 \!+\! \frac{C_0}{K_r+1}  \mathbf{w}_i^T  \mathbf{G}^T \bm{\Xi}^T \mathbf{R}_k \bm{\Xi} \mathbf{G}^* \mathbf{w}_i^* \!\right) \nonumber \\
			& - 2\Re\left\{ t_k^*  \mathbf{\bar{h}}_k^T \bm{\Xi} \mathbf{G}  \mathbf{w}_k \right\} + |t_k|^2 N_0 + 1.
		\end{align}
		
		Besides, the auxiliary variables $\mathbf{t}$, $\bm{\kappa}$ and $q$ are updated by
		\begin{equation}
			t_k \!= \!\frac{ \mathbf{\bar{h}}_k^T \bm{\Xi} \mathbf{G}  \mathbf{w}_k }{ \sum_{i=1}^K \! \left( \! \left| \mathbf{\bar{h}}_k^T \bm{\Xi} \mathbf{G} \mathbf{w}_i \right|^2 \!\!+ \!\!\frac{C_0}{K_r+1}  \mathbf{w}_i^T  \mathbf{G}^T \bm{\Xi}^T \mathbf{R}_k \bm{\Xi} \mathbf{G}^* \mathbf{w}_i^* \! \right)  \!+\! N_0},
		\end{equation}
		$\kappa_k = \epsilon_k^{-1}$, and $q = \frac{B \cdot\sum_{k=1}^K R_k }{P_{all}}$ at every iteration.	
		
		\begin{proof}
			Please see Appendix~\ref{prof_lemma1}.
		\end{proof}
	\end{lemma}
	
	Subsequently, we aim to maximize the energy efficiency by jointly optimizing $\mathbf{W}$, $\bm{\Xi}$, and $\mathbf{X}$. However, it is still difficult to solve this problem due to the coupling effects among the optimization variables. Generally, there are no standard methods to obtain a joint optimal solution for these non-convex problems. One common method is to apply the alternating optimization by iteratively optimizing $\mathbf{W}$, $\bm{\Xi}$, and $\mathbf{X}$ to solve the problem~\eqref{error1} with the other being fixed. As such, in the $j$-th iteration, we propose a Lagrange dual method to update $\mathbf{W}^{(j)}$, a multi-block manifold optimization to update $\bm{\Xi}^{(j)}$ and penalty-based method to update $\mathbf{X}^{(j)}$. The detailed update procedures are provided in the following subsections.

	\subsection{Penalty-based Alternating Optimization Algorithm}
	\subsubsection{Subproblem w.r.t. $\mathbf{W}$}
	With fixed $\bm{\Xi}$ and $\mathbf{X}$, by discarding the constant in objective function~\eqref{error1} irrelevant to $\mathbf{W}$, the optimization problem becomes
	\begin{subequations}\label{P_W1}
		\begin{align}
			\min_{\mathbf{W}} \quad & f_1(\mathbf{W}) \\
			\mathrm{s.t.} \quad & \eqref{P1_1},		
		\end{align}
	\end{subequations}
	where the objective function is given by
	\begin{equation}
		\begin{aligned}
			f_1(\mathbf{W})& =  B \cdot \sum_{i=1}^K  \mathbf{w}_i^T \sum_{k=1}^K \kappa_k |t_k|^2 \Big( \mathbf{G}^T \bm{\Xi}^T \mathbf{\bar{h}}_k \mathbf{\bar{h}}_k^H \bm{\Xi} \mathbf{G}^*  \\
			&+ \frac{C_0}{K_r + 1}  \mathbf{G}^T \bm{\Xi}^T \mathbf{R}_k \bm{\Xi}  \mathbf{G}^*  \Big)  \mathbf{w}_i^*  + \frac{q}{\nu}\sum_{k=1}^K \|\mathbf{w}_k\|^2  \\
			&- \sum_{k=1}^K 2 B \kappa_k \Re\left\{ t_k^* \mathbf{\bar{h}}_k^T \bm{\Xi} \mathbf{G}  \mathbf{w}_k \right\}.
		\end{aligned}
	\end{equation}
	In addition, the constraint~\eqref{P1_1} can be equivalently transformed as in the objective function, i.e.,
	\begin{equation}\label{P_W2}
		\min_{\mathbf{W}} \ f_1(\mathbf{W}) + \rho \left( \sum_{k=1}^K \|\mathbf{w}_k\|^2 - P_{\max}  \right),
	\end{equation} 
	where $\rho>0$ is the Lagrange multiplier. Then, the objective function of problem~\eqref{P_W2} can be rewritten as
	\begin{equation}\label{P_W3}
		\min_{\mathbf{w}} \ \mathbf{w}^T \mathbf{A}_w \mathbf{w}^* - 2 \Re\{ \bm{\zeta}_w^T \mathbf{w}\},
	\end{equation}
	where $\mathbf{w} = [\mathbf{w}_1^T,\mathbf{w}_2^T,\cdots,\mathbf{w}_K^T]^T $, $\mathbf{A}_w  = \mathrm{blkdiag}\{\mathbf{A}_{w1},$ $\mathbf{A}_{w2},\cdots,\mathbf{A}_{wK} \} $, and $\bm{\zeta}_w = [\bm{\zeta}_{w1}^T,\bm{\zeta}_{w2}^T,\cdots,\bm{\zeta}_{wK}^T]^T$. More specifically, $\mathbf{A}_{wi}$ and $\bm{\zeta}_{w1}^T$ are defined by
	\begin{align}
		\mathbf{A}_{wi} & = B \sum_{k=1}^K \kappa_k |t_k|^2 \Big(  \mathbf{G}^T \bm{\Xi}^T \mathbf{\bar{h}}_k \mathbf{\bar{h}}_k^H \bm{\Xi} \mathbf{G}^* \nonumber \\
		& \hspace{0.5cm} + \frac{C_0}{K_r + 1}  \mathbf{G}^T \bm{\Xi}^T \mathbf{R}_k \bm{\Xi} \mathbf{G}^*  \Big) + \left(\frac{q}{\nu}+ \rho\right) \mathbf{I}_{M}, \\
		\bm{\zeta}_{wk}^T & = B \kappa_k t_k^* \mathbf{\bar{h}}_k^T \bm{\Xi} \mathbf{G}.
	\end{align}
	Since $\mathbf{A}_w$ is a positive-definite matrix, the optimal $\mathbf{w}$ can be obtained by setting the first-order derivative of the objective function~\eqref{P_W3} to zero, which yields
	\begin{equation}\label{update_w}
		\mathbf{w} = (\mathbf{A}_w^T)^{-1} \bm{\zeta}_w^*,
	\end{equation}
	where the optimal $\rho$ in $\mathbf{A}_w$ is chosen according to the Karush-Kuhn-Tucker condition of the Lagrange duality transformation. It can be determined efficiently through bisection search to satisfy
	\begin{equation}\label{update_rho}
		\rho = \min \{ \rho >0:  \sum_{k=1}^K \|\mathbf{w}_{k}\|^2 \le P_{\max} \}.
	\end{equation}
	
	\subsubsection{Subproblem w.r.t. \texorpdfstring{$\bm{\Xi}$}{beta}}
	With fixed $\mathbf{W}$ and $\mathbf{X}$, by discarding the constant in objective function~\eqref{error1} irrelevant to $\bm{\Xi}$, the optimization problem becomes	
	\begin{align}\label{op_xi}
		&\min_{\hat{\bm{\xi}}} \ f_2(\hat{\bm{\xi}}) = \hat{\bm{\xi}}^T \mathbf{A}_{\xi} \hat{\bm{\xi}} - 2\sum_{k=1}^K \kappa_k \Re\big\{ t_k^* \mathbf{\bar{h}}_k^T \mathrm{diag}\left( \mathbf{G} \mathbf{w}_k \right) \hat{\bm{\xi}} \big\}  \notag\\
		&\mathrm{s.t.} \quad \eqref{P1_2},
	\end{align}	
	where $\hat{\bm{\xi}} \in \mathbb{R}^{MN \times 1}$ is composed of diagonal elements of $\bm{\Xi}$, i.e., $\bm{\Xi} = \mathrm{diag}(\hat{\bm{\xi}})$, and
	\begin{align}
		\mathbf{A}_{\xi} = \sum_{i=1}^K \sum_{k=1}^K \kappa_k |t_k|^2 \Big( \mathrm{diag}(\mathbf{G} \mathbf{w}_i) \mathbf{\bar{h}}_k \mathbf{\bar{h}}_k^H \mathrm{diag}(\mathbf{G}^* \mathbf{w}_i^*) \nonumber \\
		+ \frac{C_0}{K_r+1} \mathrm{diag}(\mathbf{G} \mathbf{w}_i) \mathbf{R}_k \mathrm{diag}(\mathbf{G}^* \mathbf{w}_i^*) \Big).
	\end{align}
	
	However, since the constraint~\eqref{P1_2} is non-convex, problem~\eqref{op_xi} cannot be solved by a standard convex optimization algorithm. We first rewrite the constraint~\eqref{P1_2} into $\|\hat{\bm{\xi}}_m\|^2 = 1$ for $\forall m$, where $\hat{\bm{\xi}}_m = [\sqrt{\xi_{m1}},\sqrt{\xi_{m2}}, \cdots, \sqrt{\xi_{mN}}]^T$. As such, the vector $\hat{\bm{\xi}}_m$ should lie on the surface of a unit sphere. Thus, this problem naturally fits into the framework of manifold optimization. Let $\mathcal{M}$ denote the Riemannian manifold used to define the constraints for $\hat{\bm{\xi}}$, which is given by $\mathcal{M} = \left\{ \hat{\bm{\xi}} \in \mathcal{R}^{MN \times 1}: \ \|\hat{\bm{\xi}}_m\|^2 = 1, \ \forall m \right\}$.
	
	It is worth noting that each block $\hat{\bm{\xi}}_m$ in vector $\hat{\bm{\xi}}$ should reside in a manifold space for $\forall m$. Hence, the update of $\hat{\bm{\xi}}$ requires computing the Riemannian gradient \cite{Riemannian} for each block individually to enforce their norm constraints inherently. Specifically, the Riemannian gradient of $f_2(\hat{\bm{\xi}})$ at $\hat{\bm{\xi}}_m$, denoted by $\bigtriangledown_{\mathcal{M}} f_2(\hat{\bm{\xi}}_m)$, can be obtained by orthogonally projecting the Euclidean gradient $\bigtriangledown f_2(\hat{\bm{\xi}}_m)$, i.e.,
	\begin{equation}\label{update_M_gra}
		\bigtriangledown_{\mathcal{M}} f_2(\hat{\bm{\xi}}_m) = \bigtriangledown f_2(\hat{\bm{\xi}}_m) - \text{tr}\left( \hat{\bm{\xi}}_m^T  \bigtriangledown \!\! f_2(\hat{\bm{\xi}}_m) \right) \hat{\bm{\xi}}_m,
	\end{equation}
	where the Euclidean gradient can be computed by $\bigtriangledown f_2(\hat{\bm{\xi}}_m) = \left[ \bigtriangledown f_2(\hat{\bm{\xi}})\right]_{(m-1)N+1:mN}$ with 
	\begin{equation}\label{update_gra}
		\bigtriangledown f_2(\hat{\bm{\xi}}) = ( \mathbf{A}_{\xi}+\mathbf{A}_{\xi}^T) \hat{\bm{\xi}} - 2 \sum_{k=1}^K \kappa_k \Re\left\{ t_k^* \mathrm{diag}\left(\mathbf{G} \mathbf{w}_k\right) \mathbf{\bar{h}}_k  \right\}.
	\end{equation}
	Finally, an additional operation is needed to map the vector $\hat{\bm{\xi}}_m$ to the manifold $\mathcal{M}$ by a retraction mapping operator, i.e.,
	
	\begin{equation}
		\hat{\bm{\xi}}_m \gets \frac{ \left( \hat{\bm{\xi}}_m-\varpi_m \bigtriangledown_{\mathcal{M}} f_2(\hat{\bm{\xi}}_m) \right)^+ }{\|\hat{\bm{\xi}}_m - \varpi_m \bigtriangledown_{\mathcal{M}} f_2(\hat{\bm{\xi}}_m)\|},
	\end{equation}
	where $(x)^+ = \max\{x,0\}$ denotes the positive-part operator applied element-wise to enforce the non-negativity of $\hat{\bm{\xi}}_m$, and $\varpi_m$ is the descent step size whose value can be updated by the Armijo backtracking line search \cite{step_size}. Then, the optimization w.r.t. $\bm{\Xi}$ is summarized in \textbf{Algorithm~\ref{algorithm_xi}}.
	
	\begin{algorithm}[t]
		\caption{Block Coordinate Manifold Optimization Algorithm for Updating $\bm{\Xi}^{(j)}$.}
		\label{algorithm_xi}
		\begin{algorithmic}[1]
			\STATE \textbf{Input:} $\hat{\bm{\xi}}^{(0)} = \mathrm{diag}^{-1}( \bm{\Xi}^{(j)}$).
			\STATE Set the inner iteration index $r=0$.
			\REPEAT
			\STATE Update $\bigtriangledown_{\mathcal{M}} f_2(\hat{\bm{\xi}}_m^{(r)})$ by~\eqref{update_M_gra} and \eqref{update_gra}.
			\STATE Update $\varpi_m^{(r+1)}$,  $\forall m$, by the Armijo backtracking line search algorithm.
			\STATE Update $\hat{\bm{\xi}}_m^{(r+1)}$ with $\frac{ \left(\hat{\bm{\xi}}_m^{(r)}-\varpi_m^{(r+1)} \bigtriangledown_{\mathcal{M}} f_2(\hat{\bm{\xi}}_m^{(r)}) \right)^+}{\|\hat{\bm{\xi}}_m^{(r)}-\varpi_m^{(r+1)} \bigtriangledown_{\mathcal{M}} f_2(\hat{\bm{\xi}}_m^{(r)})\|}$
			\STATE Update $r \gets r+1$.
			\UNTIL{the fractional decrease of $f_2(\|\hat{\bm{\xi}}\|^2)$ falls below $\epsilon_{\xi}$.}
			\STATE \textbf{Output:} $\bm{\Xi}^{(j+1)} = \mathrm{diag}(\hat{\bm{\xi}}^{(r)})$.
		\end{algorithmic}
	\end{algorithm}

	\subsubsection{Subproblem w.r.t. $\mathbf{X}$  in DSD framework}
	With fixed $\mathbf{W}$ and $\bm{\Xi}$, by discarding the constant in objective function~\eqref{error1} irrelevant to $\mathbf{X}$, the optimization problem becomes
	\begin{subequations}
		\begin{align}
			& \min_{\mathbf{X}} f_3(\mathbf{X}) \\
			& \mathrm{s.t.} \quad \eqref{P1_3},
		\end{align}
	\end{subequations}
	where the objective function is given by
	\begin{align}\label{f4}
		f_3(\mathbf{X})& = \sum_{i=1}^K \sum_{k=1}^K \kappa_k |t_k|^2 \hat{\bm{\xi}}^T \Big( \underbrace{\mathrm{diag}(\mathbf{G} \mathbf{w}_i) \mathbf{\bar{h}}_k \mathbf{\bar{h}}_k^H \mathrm{diag}(\mathbf{G}^* \mathbf{w}_i^*)}_{A_1} \nonumber \\
		&+ \frac{C_0}{K_r+1} \underbrace{\mathrm{diag}(\mathbf{G} \mathbf{w}_i) \mathbf{R}_k \mathrm{diag}(\mathbf{G}^* \mathbf{w}_i^*)}_{A_2} \Big) \hat{\bm{\xi}} \nonumber \\
		& - 2\sum_{k=1}^K \kappa_k \Re\Big\{ \underbrace{t_k^* \mathbf{\bar{h}}_k^T \mathrm{diag}\left( \mathbf{G} \mathbf{w}_k \right)}_{A_3} \hat{\bm{\xi}} \Big\}.
	\end{align}
	
	Based on the proposed DSD framework in Section~\ref{sec:DSP}, we introduce two auxiliary variables to represent the amplitude and phase components of the objective function~\eqref{f4} as
	\begin{align}\label{bmnk}
		u_{mnk} & = \exp(-\alpha_g x_{mn}) r_{mnk}^{-\frac{\beta_u}{2}}, \\
		\label{cmnk}
		z_{mnk} & = \exp\left( -j\frac{2\pi}{\lambda_g} x_{mn} - j \frac{2\pi}{\lambda} r_{mnk} \right).
	\end{align}
	Then, the objective function~\eqref{f4} can be rewritten through
	\begin{align}
		A_1 & =  \frac{C_0 K_r}{K_r+1} \mathrm{diag}(\mathbf{\bar{w}}_i) (\mathbf{u}_k \circ \mathbf{z}_k) (\mathbf{u}_k \circ \mathbf{z}_k)^H \mathrm{diag}(\mathbf{\bar{w}}_i^*), \\
		A_2 & =  \mathrm{diag}(\mathbf{\bar{w}}_i) \mathrm{diag}(\mathbf{u}_k \circ \mathbf{u}_k^*) \mathrm{diag}(\mathbf{\bar{w}}_i^*), \\
		A_3 & =  t_k^* \sqrt{\frac{C_0 K_r}{K_r+1}}(\mathbf{u}_k \circ \mathbf{z}_k)^T \mathrm{diag}(\mathbf{\bar{w}}_k),
	\end{align}
	where $\mathbf{\bar{w}}_i = [w_{1i} \mathbf{1}_N^T, w_{2i} \mathbf{1}_N^T,\cdots, w_{Mi} \mathbf{1}_N^T]^T$, $\mathbf{u}_k = [u_{11k},\!\cdots\!,u_{1Nk},\!\cdots\!,u_{m1k},u_{m2k},\!\cdots\!,u_{mNk}, \!\cdots\!,u_{MNk}]^T$ and $\mathbf{z}_k =[z_{11k},\!\cdots\!,z_{1Nk},\!\cdots\!,z_{m1k},z_{m2k}, \!\cdots\!,	z_{mNk},\!\cdots\!,z_{MNk}]^T $.

	Thus, the optimization problem w.r.t. $\mathbf{X}$ can be transformed into the problem of optimizing two independent vectors, $\mathbf{u}$ and $\mathbf{z}$, through the proposed DSD design, as
	\begin{subequations}\label{op_b_c}
		\begin{align}
			&\min_{\mathbf{\mathbf{u},\mathbf{z}}} \ f_4(\mathbf{u},\mathbf{z})\\
			&\mathrm{s.t.} \ \|z_{mnk}\| = 1,\ \text{for} \ \forall m, n,k, \label{f4_1}\\
			& \qquad u_{mnk} = \exp(-\alpha_g x_{mn})  r_{mnk}^{-\frac{\beta_u}{2}} ,\ \text{for} \ \forall m, n,k, \label{f4_2} \\
			& \qquad \eqref{P1_3},
		\end{align}
	\end{subequations}
	where $\mathbf{u} = [\mathbf{u}_1^T, \mathbf{u}_2^T, \cdots, \mathbf{u}_K^T]^T $, $\mathbf{z} = [\mathbf{z}_1^T, \mathbf{z}_2^T, \cdots, \mathbf{z}_K^T]^T$ and 
	\begin{align}
		f_4(\mathbf{u},& \mathbf{z})= \sum_{i=1}^K \sum_{k=1}^K \kappa_k |t_k|^2 \hat{\bm{\xi}}^T \Big(\frac{C_0 K_r}{K_r+1} \mathrm{diag}(\mathbf{\bar{w}}_i)  \notag\\
		&\times (\mathbf{u}_k \circ \mathbf{z}_k)(\mathbf{u}_k \circ \mathbf{z}_k)^H \mathrm{diag}(\mathbf{\bar{w}}_i^*) \notag\\
		&+ \frac{C_0}{K_r+1} \mathrm{diag}(\mathbf{\bar{w}}_i) \mathrm{diag}(\mathbf{u}_k \circ \mathbf{u}_k^*) \mathrm{diag}(\mathbf{\bar{w}}_i^*)\Big) \hat{\bm{\xi}}\notag \\
		& - 2\sum_{k=1}^K \kappa_k \Re\big\{ t_k^* \sqrt{\frac{C_0 K_r}{K_r+1}}(\mathbf{u}_k \circ \mathbf{z}_k)^T \mathrm{diag}(\mathbf{\bar{w}}_k) \hat{\bm{\xi}} \big\}.
	\end{align}

	The constraints~\eqref{f4_1} and \eqref{f4_2} are both non-convex in $\mathbf{u}$ and $\mathbf{z}$, which hinders the efficient solution of problem~\eqref{op_b_c}. Hence, we introduce two auxiliary vectors $\mathbf{b}$ and $\mathbf{c}$ with the equalities $\mathbf{z} = \mathbf{c}$ and $\mathbf{u} = \mathbf{b}$. As such, problem~\eqref{op_b_c} can be rewritten by replacing $\mathbf{u}$ and $\mathbf{z}$ with $\mathbf{b}$ and $\mathbf{c}$, respectively, yielding
	\begin{subequations}\label{admm1}\vspace{2mm}
		\begin{align}
			\min_{\mathbf{b},\mathbf{c},\mathbf{u},\mathbf{z}} \  \quad &f_4(\mathbf{b},\mathbf{c}) \\
			\mathrm{s.t.} \quad & \mathbf{u} =  \mathbf{b}, \\
			& \mathbf{z} = \mathbf{c}, \\
			& \eqref{P1_3}, \eqref{f4_2}, \eqref{f4_1}.
		\end{align}
	\end{subequations}
	The augmented Lagrangian function of problem~\eqref{admm1} can incorporate equality constraints into the objective function by introducing corresponding penalty terms\cite{pdd}, written as\vspace{2mm}
	\begin{equation}
		\mathcal{L}(\mathbf{b},\mathbf{u}, \mathbf{c},\mathbf{z}) = f_4(\mathbf{b},\mathbf{c})+ \varrho \left( \|\mathbf{u}-\mathbf{b}\|^2 + \|\mathbf{z}-\mathbf{c}\|^2 \right),
	\end{equation}
	where $\varrho > 0$ is the quadratic penalty parameter of the equality constraints and updated in the outer iteration. Let $( \mathbf{u}^{(0)}, \mathbf{b}^{(0)}, \mathbf{z}^{(0)}, \mathbf{c}^{(0)})$ be initial variables as feasible values. The proposed penalty-based method iteratively performs the following steps as\vspace{2mm}
	\begin{equation}\label{admm}
		\left\{
		\begin{array}{@{}l@{}}
			\mathbf{b}^{(j+1)} = \argmin\limits_{\mathbf b}\,
			\mathcal{L}\!\left(\mathbf b,\mathbf u^{(j)},\mathbf c^{(j)},\mathbf z^{(j)}\right),\\
			\mathbf{u}^{(j+1)} = \argmin\limits_{\mathbf u,\eqref{P1_3},\eqref{f4_2}}\,
			\mathcal{L}\!\left(\mathbf b^{(j+1)},\mathbf u,\mathbf c^{(j)},\mathbf z^{(j)}\right),\\
			\mathbf{c}^{(j+1)} = \argmin\limits_{\mathbf c}\,
			\mathcal{L}\!\left(\mathbf b^{(j+1)},\mathbf u^{(j+1)},\mathbf c,\mathbf z^{(j)}\right),\\
			\mathbf{z}^{(j+1)} = \argmin\limits_{\mathbf z,\eqref{P1_3},\eqref{f4_1}}\,
			\mathcal{L}\!\left(\mathbf b^{(j+1)},\mathbf u^{(j+1)},\mathbf c^{(j+1)},\mathbf z\right).
		\end{array}
		\right.
	\end{equation}\vspace{2mm}
	Then, we will divide in into two subproblems for obtaining $\{\mathbf{b}^{(j+1)}, \mathbf{u}^{(j+1)} \}$ and $\{  \mathbf{c}^{(j+1)},\mathbf{z}^{(j+1)}\}$.
	
	$\bullet$ \emph{Optimization of $\{\mathbf{b}^{(j+1)}, \mathbf{u}^{(j+1)} \}$:}
	By dropping irrelevant variables, the optimization problem of this block becomes\vspace{2mm}
	\begin{equation}\label{op_b_u}
		\begin{aligned}
			&\min_{\mathbf{b},\mathbf{u}} \ \mathbf{b}^T \mathbf{A}_b \mathbf{b} - 2  \Re\left\{ \bm{\zeta}_b^T \mathbf{b} \right\} + \varrho  \|\mathbf{u}-\mathbf{b}\|^2  \\
			& \mathrm{s.t.} \quad \eqref{P1_3},  \ \eqref{f4_2}, 
		\end{aligned}
	\end{equation}
	where $\mathbf{A}_b  = \mathrm{blkdiag}\{ 	\mathbf{A}_{b1},	\mathbf{A}_{b2},\cdots,	\mathbf{A}_{bK} \}$,  $\bm{\zeta}_b = [\bm{\zeta}_{b1}^T,\bm{\zeta}_{b2}^T,\cdots,\bm{\zeta}_{bK}^T]^T $, $\bm{\zeta}_{bk} = \kappa_k t_k^* \mathrm{diag}(\mathbf{c}_k) \mathrm{diag}(\mathbf{\bar{w}}_k) \hat{\bm{\xi}}$,
	and\vspace{2mm}
	\begin{align}
		\mathbf{A}_{bk}= \sum_{i=1}^K \kappa_k |t_k|^2 \mathrm{diag}(\hat{\bm{\xi}}) \mathrm{diag}(\mathbf{\bar{w}}_i)  \Bigg(\frac{C_0 K_r}{K_r+1} \mathbf{c}_k \mathbf{c}_k^H \nonumber \\
		+ \frac{C_0}{K_r+1} \mathbf{I}_{MN}\Bigg) \mathrm{diag}(\mathbf{\bar{w}}_i^*) \mathrm{diag}(\hat{\bm{\xi}}).
	\end{align}
	It can be observed that the optimization problem~\eqref{op_b_u} w.r.t. $\mathbf{b}$ is an unconstrained least-squares problem, which can be solved optimally by setting its first-order derivative to zero and yielding\vspace{2mm}
	\begin{equation}\label{update_b}
		\mathbf{b} = \left( \mathbf{A}_b + \mathbf{A}_b^T + 2\varrho \mathbf{I}_{MNK} \right)^{-1} \left(  2\bm{\zeta}_b^*  + 2\varrho \mathbf{u}  \right).
	\end{equation}
	The update of $\mathbf{u}$ should satisfy the constraint~\eqref{f4_2} and minimize the equality penalty term of the objective function~\eqref{op_b_u}. This can be efficiently solved through a one-dimensional search over the coarse grid set $\mathbf{x}_{m}^{CG}$. In particular, the design of $\mathbf{x}_{m}^{CG}$ ensures that the grid values lie within $[x_m^{\min},x_m^{\max}]$ and that any two grid points are separated by at least $\Delta x$, thereby naturally satisfying constraint~\eqref{P1_3}. For $\forall m, n$, a one-dimensional search for coarse grids is performed by
	\begin{equation}\label{update_coarse}
		\begin{aligned}
			x_{mn}^C= \argmin_{x_{mn} \in \mathbf{x}_{m}^{CG}} \sum_{k=1}^K	\bigg| [\mathbf{b}]_{mnk} -  \exp(-\alpha_g x_{mn}) \\
			\times \left[(x_k^U - x_{mn})^2+(y_k^U - \bar{Y}_m)^2\right]^{-\frac{\beta_u}{4}}\bigg|^2,
		\end{aligned}
	\end{equation} 
	where $x_{mn}^C \neq x_{mn'}^C$, for $\forall n,n'$. Then, the update of $\mathbf{u}$ is given for $\forall m,n,k$ by
	\begin{align}\label{update_u}
		[\mathbf{u}]_{mnk} = & \exp\left(-\alpha_g x_{mn}^C\right) \nonumber \\
		& \times \left[(x_k^U - x_{mn}^C)^2+(y_k^U - \bar{Y}_m)^2\right]^{-\frac{\beta_u}{4}}.
	\end{align}
	
	\vspace{4mm}
	$\bullet$ \emph{Optimization of $\{  \mathbf{c}^{(j+1)},\mathbf{z}^{(j+1)}\}$:} By dropping irrelevant variables, the optimization problem of this block becomes
	\begin{subequations}\label{block2}
		\begin{align}
			\min_{\mathbf{c},\mathbf{z}} \quad & \mathbf{c}^T \mathbf{A}_c \mathbf{c}^* - 2 \Re\left\{  \bm{\zeta}_c^T \mathbf{c} \right\} + \varrho  \|\mathbf{z}-\mathbf{c}\|^2  \\
			\mathrm{s.t.} \quad & \eqref{f4_1},
		\end{align}
	\end{subequations}
	where $\mathbf{A}_c = \mathrm{blkdiag}\{\mathbf{A}_{c1},\mathbf{A}_{c2},\cdots,\mathbf{A}_{cK}  \}$,
	$\bm{\zeta}_c = [\bm{\zeta}_{c1}^T,\bm{\zeta}_{c2}^T,\cdots,\bm{\zeta}_{cK}^T]^T $, $\bm{\zeta}_{ck} = \kappa_k t_k^* \mathrm{diag}(\mathbf{b}_k) \mathrm{diag}(\mathbf{\bar{w}}_k) \hat{\bm{\xi}}$,
	and
	\begin{align}
		\mathbf{A}_{ck} = \frac{C_0 K_r}{K_r+1} \sum_{i=1}^K \kappa_k |t_k|^2 \mathrm{diag}(\hat{\bm{\xi}}) \mathrm{diag}(\mathbf{\bar{w}}_i)   \mathbf{b}_k \nonumber \\
		\times \mathbf{b}_k^H  \mathrm{diag}(\mathbf{\bar{w}}_i^*) \mathrm{diag}(\hat{\bm{\xi}}).
	\end{align}
	Similarly, the optimization~\eqref{block2} w.r.t. $\mathbf{c}$ is an unconstrained least-squares problem, which can be solved optimally by setting its first-order derivative to zero and yielding
	\begin{equation}\label{update_c}
		\mathbf{c}^{(j+1)} = \left( \mathbf{A}_c^T + \varrho \mathbf{I}_{MNK} \right)^{-1} \left(  \bm{\zeta}_c^* + \varrho \mathbf{z}^{(j)}  \right).
	\end{equation}
	The update of $\mathbf{z}$ should satisfy the constraint~\eqref{f4_1} and minimize the equality penalty term of the objective function~\eqref{block2}. This can also be efficiently solved through a one-dimensional search over the fine grid set $\mathbf{x}_{m}^{FG}$. For $\forall m, n$, a one-dimensional search for fine grids is performed by
	\begin{align}\label{update_fine}
		\Delta x_{mn}^F \!=\! \argmin_{\Delta x_{mn} \in \mathbf{x}_{m}^{FG}} \! \sum_{k=1}^K \!	\Bigg| [\mathbf{c}]_{mnk} \!-\!  \exp\!\Bigg(\!\! -\!j\frac{2\pi}{\lambda_g} (x_{mn}^C\! +\! \Delta x_{mn}) \nonumber \\
		- j \frac{2\pi}{\lambda} \sqrt{(x_k^U - (x_{mn}^C + \Delta x_{mn}))^2+(y_k^U - \bar{Y}_m)^2} \Bigg) \Bigg|^2.
	\end{align} 
	Then, the update of $\mathbf{z}$ is given for $\forall m,n,k$ by
	\begin{align}\label{update_z}
		& [\mathbf{z}]_{mnk} = \exp\Big(-j\frac{2\pi}{\lambda_g} (x_{mn}^C + \Delta x_{mn}^F) \nonumber\\
		& \hspace{0.3cm} - j \frac{2\pi}{\lambda} \sqrt{(x_k^U - (x_{mn}^C + \Delta x_{mn}^F))^2+(y_k^U - \bar{Y}_m)^2} \Big).
	\end{align}
	
	\vspace{2mm}
	\begin{remark}\label{remark:complexity}
		In the proposed DSD framework, the dual-scale approach performs two sequential one-dimensional searches with total complexity on the order $\mathcal{O}(G_C + G_F)$, where $G_C$ and $G_F$ denote the numbers of coarse-grid and fine-grid search points, respectively. It substantially reduces the computational complexity compared with prior methods whose joint search \cite{model_PA} has complexity on the order of $\mathcal{O}(G_C G_F)$.
	\end{remark}
	
	The overall algorithm for joint transmit precoding, PA power radiation, and PA deployment is summarized in \textbf{Algorithm~2}. The four implementation protocols detailed in Section~\ref{sec:DSP} can all be readily accommodated in \textbf{Algorithm~2}. More particularly, the analytical expressions of $\mathbf{x}_m^{CG}$ and $\mathbf{x}_m^{FG}$ for protocols~(a)-(d) can be obtained from Eqs.~\eqref{dsp1} to \eqref{dsp2} with well-designed $\delta^F$ and $\delta^C$.
	
	\begin{algorithm}[t]
		\caption{Proposed Alternating Optimization Algorithm for Joint BS Transmit Precoding, PA Power Radiation, and Dual-Scale PA Positioning.}
		\label{algorithm_all}
		\begin{algorithmic}[1]
			\STATE  Initialize equal power radiation and equal-interval position deployment for PAs, i.e., $\bm{\Xi}^{(0)} = \sqrt{\frac{1}{N}} \mathbf{I}_{MN}$ and $x_{mn}^{(0)} = \frac{x_m^{\max} - x_m^{\min}}{N+1}n$. Initialize the precoding matrix $\mathbf{W}^{(0)}$ by MRT scheme. Initialize $\rho^{(0)}=0$.
			\REPEAT
			\STATE Set the iteration index $j=0$.
			\REPEAT 
			\STATE Update $t_k^{(j+1)}$, $\kappa_k^{(j+1)}$, for $\forall k$, and $q^{(j+1)}$ by \textbf{Lemma~1}.			
			\STATE Update $\mathbf{W}^{(j+1)}$ as~\eqref{update_w}.
			\STATE Update $\rho^{(j+1)}$ as~\eqref{update_rho}.
			\STATE Update $\bm{\Xi}^{(j+1)}$ by \textbf{Algorithm 1}.
			\STATE Update $\mathbf{b}^{(j+1)}$ and $\mathbf{u}^{(j+1)}$ as~\eqref{update_b} and \eqref{update_u}, respectively.
			\STATE Update $\mathbf{c}^{(j+1)}$ and $\mathbf{z}^{(j+1)}$ as Eqs.~\eqref{update_c} and \eqref{update_z}, respectively.
			\STATE Update $x_{mn}^{(j+1)} = x_{mn}^C + \Delta x_{mn}^F$ by Eqs.~\eqref{update_coarse} and \eqref{update_fine}, for $\forall m,n$.
			\STATE Update $j \gets j+1$.
			\UNTIL{the fractional decrease of the objective value of $\eta_{\text{EE}}$ falls below a predefined threshold $\epsilon_{\text{in}}$}
			\STATE Update the penalty factor $\varrho \gets  c_{\varrho} \cdot \varrho$.
			\UNTIL{the fractional decrease of the objective value of $\eta_{\text{\text{EE}}}$ falls below a predefined threshold $\epsilon_{\text{out}}$}
			\STATE \textbf{Output:} The BS transmit precoding $\mathbf{W}^{(j+1)}$, PA radiation power allocation $\bm{\Xi}^{(j+1)}$, PA position $\mathbf{X}^{(j+1)}$.
		\end{algorithmic}
	\end{algorithm}

\subsection{Convergence Analysis}
For a given penalty factor $\varrho>0$, define the augmented objective in the inner loop of Algorithm~2 as
\begin{align}\label{eq:aug_obj_final}
\mathcal{F}(\mathbf{W},\bm{\Xi},\mathbf{b},\mathbf{u},\mathbf{c},\mathbf{z};\mathbf{t},\bm{\kappa},q)
&\triangleq B\sum_{k=1}^{K}\kappa_{k}\epsilon_{k}+qP_{\mathrm{all}} \nonumber\\
&\quad+\varrho\!\left(\|\mathbf{u}-\mathbf{b}\|^{2}+\|\mathbf{z}-\mathbf{c}\|^{2}\right).
\end{align}
The inner loop generates a non-increasing sequence of $\mathcal{F}$. The updates of $\{t_k\}$, $\{\kappa_k\}$, and $q$ follow Lemma~\ref{lemma1} and yield the equivalent reformulation in \eqref{error1}. For fixed $(\bm{\Xi},\mathbf{X})$, the $\mathbf{W}$-subproblem in \eqref{P_W3} is a strictly convex quadratic problem since $\mathbf{A}_w\succ\mathbf{0}$, and \eqref{update_w} with \eqref{update_rho} yields the update of $\mathbf{W}$. For fixed $(\mathbf{W},\mathbf{X})$, the $\bm{\Xi}$-subproblem is handled by Riemannian gradient descent with Armijo backtracking, which decreases the objective value at each iteration. For fixed $(\mathbf{W},\bm{\Xi})$, the updates of $\mathbf{b}$ and $\mathbf{c}$ in \eqref{update_b} and \eqref{update_c} are obtained in closed form, while the updates of $\mathbf{u}$ and $\mathbf{z}$ in \eqref{update_u} and \eqref{update_z} are obtained by searches over the feasible coarse-grid and fine-grid sets. Hence, $\mathcal{F}^{(j+1)}\le \mathcal{F}^{(j)},\ \forall j$.

Since the feasible set defined by \eqref{P1_1}--\eqref{P1_3} is bounded, $\mathcal{F}$ is lower bounded. Thus, for any fixed $\varrho>0$, the inner loop converges in penalized objective value. In the outer loop, the penalty factor is updated as $\varrho\leftarrow c_{\varrho}\varrho$ with $c_{\varrho}>1$. As $\varrho$ increases, the penalty terms $\|\mathbf{u}-\mathbf{b}\|^{2}$ and $\|\mathbf{z}-\mathbf{c}\|^{2}$ are progressively reduced. Therefore, Algorithm~2 converges in penalized objective value, while the penalty terms are progressively reduced.

\subsection{Computational Complexity Analysis}\label{sec:complexity}

The computational complexity of Algorithm~2 is characterized by the costs of the block updates in each inner iteration.

\begin{itemize}
\item \emph{Update of $\{t_k,\kappa_k,q\}$:} $\mathcal{O}(K^2MN)$.

\item \emph{Update of $\mathbf{W}$:} Solving \eqref{update_w} with the bisection search in \eqref{update_rho} requires complexity $\mathcal{O}(I_{\rho}KM^3)$, where $I_{\rho}$ denotes the number of bisection steps.

\item \emph{Update of $\bm{\Xi}$:} The manifold optimization in Algorithm~\ref{algorithm_xi} requires complexity $\mathcal{O}(I_{\xi}I_aM^2N^2)$, where $I_{\xi}$ and $I_a$ denote the numbers of Riemannian and backtracking iterations, respectively.

\item \emph{Update of $\{\mathbf{b},\mathbf{u}\}$:} The update of $\mathbf{b}$ in \eqref{update_b} requires $\mathcal{O}(K(MN)^3)$, and the coarse-grid search for $\mathbf{u}$ in \eqref{update_coarse} requires $\mathcal{O}(MNKG_C)$.

\item \emph{Update of $\{\mathbf{c},\mathbf{z}\}$:} The update of $\mathbf{c}$ in \eqref{update_c} requires $\mathcal{O}(K(MN)^3)$, and the fine-grid search for $\mathbf{z}$ in \eqref{update_fine} requires $\mathcal{O}(MNKG_F)$.
\end{itemize}

Therefore, the overall computational complexity is $\mathcal{O}\Big(
I_{\mathrm{out}}I_{\mathrm{in}}
\big[
K^2MN+I_{\rho}KM^3+I_{\xi}I_aM^2N^2+K(MN)^3+MNK(G_C+G_F)
\big]
\Big)$, where $I_{\mathrm{in}}$ and $I_{\mathrm{out}}$ denote the numbers of inner and outer iterations in Algorithm~2, respectively. As indicated in Remark~\ref{remark:complexity}, the DSD-based deployment update only contributes a linear search term $MNK(G_C+G_F)$ to the overall complexity. In addition, the block-coordinate structure yields closed-form updates for $\mathbf{b}$ and $\mathbf{c}$, while confining the updates of $\mathbf{u}$ and $\mathbf{z}$ to structured feasible sets. These features make Algorithm~2 a low-complexity approach for the considered joint design problem.

	\section{Numerical Results}\label{sec:simulation}
	In this section, we will demonstrate the advantages of PASS and the effectiveness of the proposed algorithms in the proposed DSD framework. In particular, we consider the scenario over a square dense urban area of size $100 \times 100$ m${}^2$. The waveguides are uniformly arranged to serve randomly distributed users. Then, the position of deployed PA on the $m$-th waveguide can be expressed as $\bar{Y}_m = \frac{100}{M+1}m$, $x_m^{\min} = 0$ and $x_m^{\max} = 100$, for $\forall m$. Unless stated otherwise, the simulation setup is given in Table~\ref{simu}.
	
	\renewcommand{\arraystretch}{1}
	
	\begin{table}[t!]
		\centering
		\begin{tabular}{>{\centering\arraybackslash}m{3.2cm}|>{\centering\arraybackslash}m{4.6cm}}
			\hline
			\textbf{Parameters} & \textbf{Values} \\
			\hline 	
			Number of waveguides & $M=3$ \\ \hline	
			PA number per waveguide & $N=4$ \\ \hline
			Transmission bandwidth & $B = 180$ MHz \\ \hline
			Number of users & $K=4$ \\ \hline
			Position of users & $\mathbf{p}_1^U = [15.9 , 54.3]$, $\mathbf{p}_2^U = [98.6, 85.4]$, $\mathbf{p}_3^U = [74.5, 24.1]$, $\mathbf{p}_4^U = [37.4, 23.9]$ \\ \hline
			Carrier frequency and wavelength & $f_c = 28$ GHz, $\lambda = \frac{c}{f_c}=1.07 \times 10^{-2}$ m, $\lambda_g = \frac{\lambda}{1.4}=7.65 \times 10^{-3}$ m \\
			\hline
			Rician factor & $K_r = 0.5$\\ \hline
			In-waveguide attenuation & $\alpha_g = -18$ dB\\ \hline
			Wireless attenuation & $C_0=\left(\frac{\lambda}{4\pi}\right)^2 = -61.4$ dB, $\beta_u = 2.2$\\ \hline
			Transmit power efficiency & $\nu=0.9$\\ \hline
			Each PA activation power consumption & $P_{\text{PA}}^{\text{act}} = 5$ dBm\\ \hline
			Each PA motorized module power consumption & $P_{\text{PA}}^{\text{mot}} = 20$ dBm\\ \hline
			Each PA piezoelectric module power consumption & $P_{\text{PA}}^{\text{pie}} = 8$ dBm\\ \hline
			Static BS power consumption & $P_{\text{BS}}^{\text{sta}} = 9$ dB\\ \hline
			Noise power & $N_0 = -80$ dBm\\ \hline
			Initial step size & $\varrho_m^{(0)}=10^2$\\ \hline
			Predefined thresholds & $\epsilon_{\xi}=10^{-4}$, $\epsilon_{\text{in}}=\epsilon_{\text{out}}=10^{-8}$ \\ \hline
			PA's sliding length & $L^C = 0.2$ m \\ \hline
			Positioning accuracy of PBs and PAs & $\delta^C = 1$ m, $\delta^F = 10^{-4}$ m \\ \hline			
		\end{tabular}
		\caption{Simulation parameters and values.}
		\label{simu}
	\end{table}
	
	\subsection{Validation of Derived Theoretical Results}
\begin{figure}[t]
		\centering
		\includegraphics[width=0.5\textwidth]{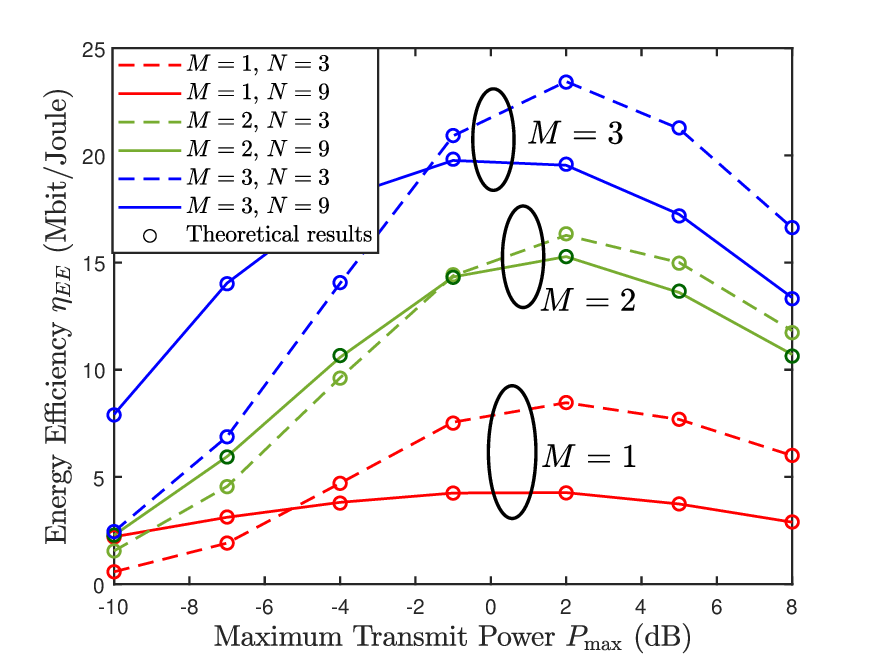}
		\caption{Theoretical results of energy efficiency with Monte Carlo results.}
		\label{fig_EE_simulation}
	\end{figure}

	Fig.~\ref{fig_EE_simulation} validates the theoretical results of energy efficiency given in~\eqref{eq_EE}, where the transmit precoding $\mathbf{W}$ is obtained by the maximum ratio transmission scheme \cite{mrt}, and both the PAs' radiation power allocation and their positions are uniformly generated. Across all $P_{\max}$, $M$, and $N$, the theoretical results, indicated by circular markers, closely overlap with the Monte Carlo curves, thereby confirming the accuracy of the derived energy efficiency expression. It can also be observed that increasing $M$ yields significant performance gains, whereas continuously increasing $P_{\max}$ and $N$ degrades energy efficiency. This arises from the fact that adding more waveguides enhances the spatial diversity gain, while increasing the transmit power or the PA number, in this non-optimized setting, brings only marginal spectral-efficiency improvements but substantially higher power consumption. These observations highlight the necessity of jointly designing transmit precoding, PA radiation power allocation, and PA deployment to fully realize the energy efficiency potential of PASS.
	
	\subsection{Convergence Analysis of the Proposed Algorithm}
	
	\begin{figure}[t]
		\centering
		\subfigure[Convergence behavior of $\hat{\bm{\xi}}$]{
			\includegraphics[width=0.5\textwidth]{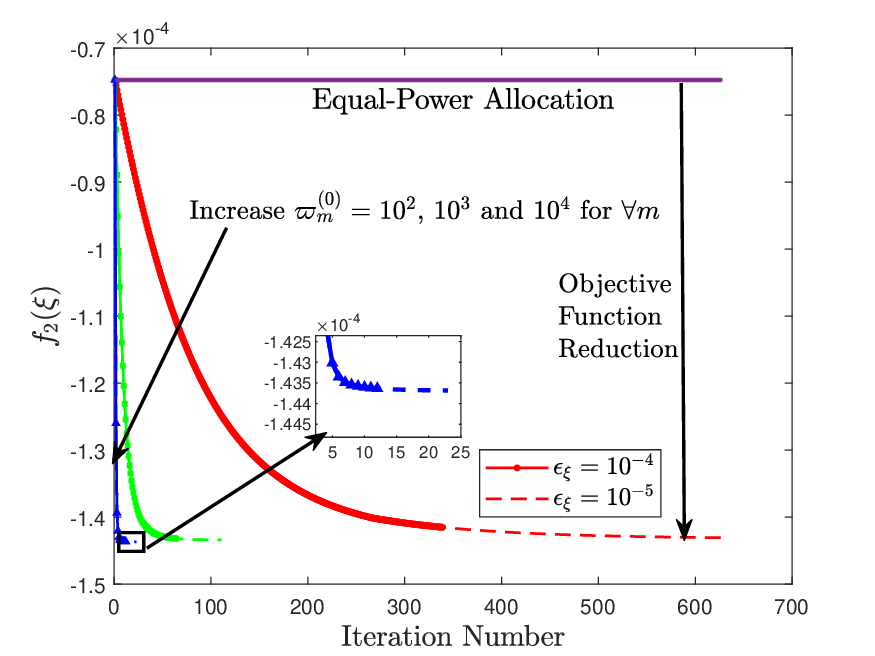}  
			\label{fig_coverage_xi}
		}
		\hfill
		\subfigure[Convergence behavior of $\eta_{EE}$]{
			\includegraphics[width=0.5\textwidth]{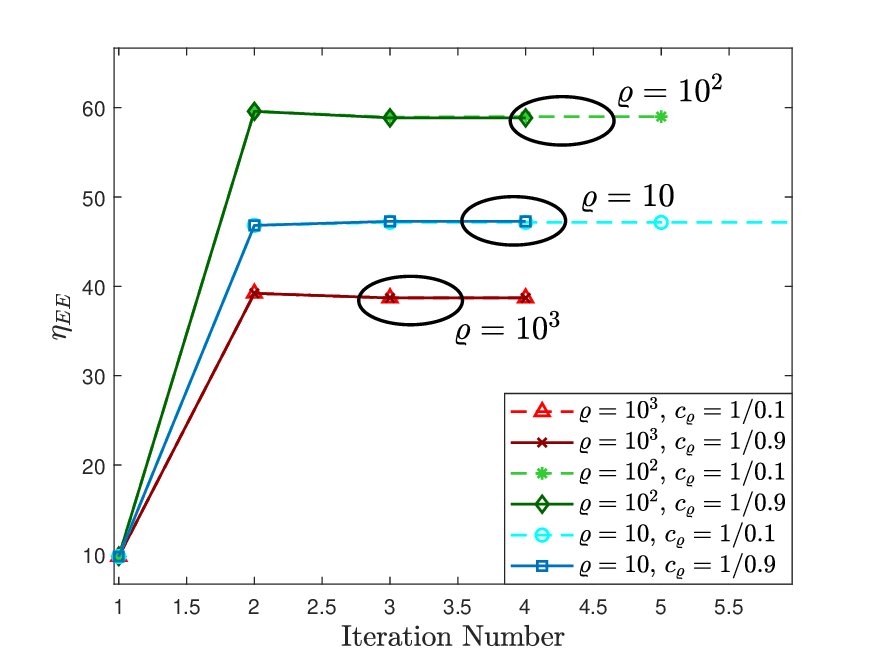} 
			\label{fig_coverage_EE}
		}
		\caption{Convergence behavior of Algorithm 1 and 2.} 
		\label{fig_coverage}
	\end{figure}
	
	Fig.~\ref{fig_coverage} demonstrates the convergence behavior of the proposed algorithm. In particular, Fig.~\ref{fig_coverage_xi} verifies the convergence of \textbf{Algorithm 1} within the inner iteration loop. It shows that under different settings of $\bm{\varpi}^{(0)}$ and $\epsilon_{\xi}$, the objective function $f_2(\hat{\bm{\xi}})$ converges to the same value and all achieve substantial performance gains over equal-power allocation. Besides, with a proper choice of initial step size to $10^2$, the iteration number can be reduced by nearly $60$-fold. Moreover, it is observed in Fig.~\ref{fig_coverage_EE} that after the first iteration of \textbf{Algorithm 2}, the energy efficiency already reaches the same level as the final performance, and converges within only a few subsequent iterations.

	\subsection{Energy Efficiency Comparison: PASS vs. Conventional MIMO and Cell-Free Systems}
	\begin{figure}[htbp]
		\centering
		\includegraphics[width=0.5\textwidth]{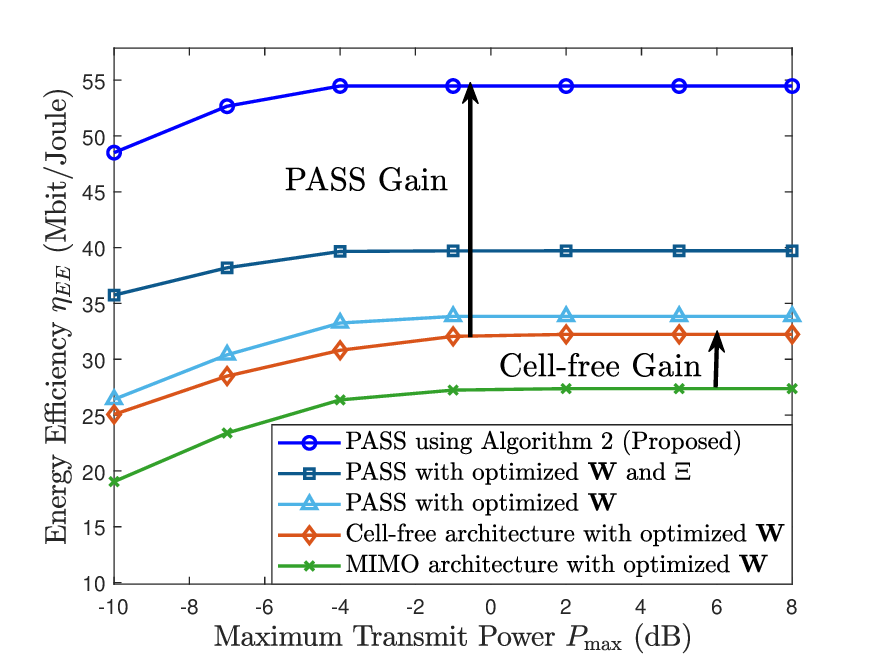}
		\caption{Comparison with MIMO and cell-free architectures, for $\varrho=230$ and $c_{\varrho}=1/0.9$.}
		\label{fig_EE_compare_cellfree}
	\end{figure}
	
	Fig.~\ref{fig_EE_compare_cellfree} considers conventional MIMO and cell-free architectures as performance benchmarks. For a fair comparison, it is assumed that these two architectures employ the same number $M$ of RF chains and transmit precoding scheme as in PASS. Due to space limitations, we only present the received signal of the $k$-th user for the MIMO and cell-free as
	\begin{align}
		y_k^{\text{MIMO}} & = (\mathbf{h}_k^{\text{MIMO}})^T \sum_{i=1}^K \mathbf{w}_i^{\text{MIMO}} s_i + n_k, \\
		y_k^{\text{cell-free}} & = \sum_{m=1}^M (\mathbf{h}_{mk}^{\text{cell-free}})^T \sum_{i=1}^K \mathbf{w}_{mi}^{\text{cell-free}} s_i + n_k
	\end{align}
	where $\mathbf{h}_k^{\text{MIMO}} \in \mathbb{C}^{M \times 1}$ and $\mathbf{h}_{mk}^{\text{cell-free}} \in \mathbb{C}^{N \times 1}$ are the channel from the $k$-th user to the MIMO-BS and to the $m$-th cell-free-BS, respectively. The MIMO-BS is positioned at $(0,0)$, while the $m$-th cell-free-BS is positioned at $ \left(50, \frac{100}{M+1}m\right)$. The corresponding energy efficiencies are
	\begin{equation}
		\eta_{\mathrm{EE}}^{\mathrm{MIMO}} = \frac{B \sum_{k=1}^K R_k^{\mathrm{MIMO}}}{P_{\mathrm{all}}^{\mathrm{MIMO}}}, \quad
		\eta_{\mathrm{EE}}^{\mathrm{cell-free}} = \frac{B \sum_{k=1}^K R_k^{\mathrm{cell-free}}}{P_{\mathrm{all}}^{\mathrm{cell-free}}},
	\end{equation}
	 where the achievable rates are given by $R_k^{\mathrm{MIMO}} = \log_2\!\Big(1+\frac{\left|(\mathbf{h}_k^{\mathrm{MIMO}})^T\mathbf{w}_k^{\mathrm{MIMO}}\right|^2}{\sum_{i\neq k}^K \left|(\mathbf{h}_k^{\mathrm{MIMO}})^T\mathbf{w}_i^{\mathrm{MIMO}}\right|^2+N_0}\Big)$ and $R_k^{\mathrm{cell-free}} = \log_2\!\Big(1+\frac{\left|\sum_{m=1}^M (\mathbf{h}_{mk}^{\mathrm{cell-free}})^T \mathbf{w}_{mk}^{\mathrm{cell-free}}\right|^2}{\sum_{i\neq k}^K \left|\sum_{m=1}^M (\mathbf{h}_{mk}^{\mathrm{cell-free}})^T \mathbf{w}_{mi}^{\mathrm{cell-free}}\right|^2+N_0}\Big)$. The corresponding power consumptions are $P_{\mathrm{all}}^{\mathrm{MIMO}} = \sum_{k=1}^K \frac{\|\mathbf{w}_k^{\mathrm{MIMO}}\|^2}{\nu} + P_{\text{BS}}^{\text{sta}}$ and $P_{\mathrm{all}}^{\mathrm{cell-free}} = \sum_{m=1}^M \sum_{k=1}^K \frac{\|\mathbf{w}_{mk}^{\mathrm{cell-free}}\|^2}{\nu} + M P_{\text{BS}}^{\text{sta}}$.
	
	It can be observed from Fig.~\ref{fig_EE_compare_cellfree} that, as $P_{\max}$ increases, all the energy-efficiency curves first rise and then approach saturation. Specifically, at $P_{\max}=5$ dB, the cell-free architecture achieves $17.7\%$ higher energy efficiency than conventional MIMO by virtue of the spatial-diversity gain from the cooperation of $M$ BSs. Because PASS adopts a more distributed architecture, the third curve in Fig.~\ref{fig_EE_compare_cellfree} already exhibits performance gains under the same configuration. By further optimizing the PA radiation power allocation and positions, a larger energy-efficiency gain is achieved, yielding an additional improvement of about $70\%$ with respect to the cell-free architecture.

	\subsection{Impact of Waveguide and PA Numbers}
	\begin{figure}[t]
		\centering
		\includegraphics[width=0.5\textwidth]{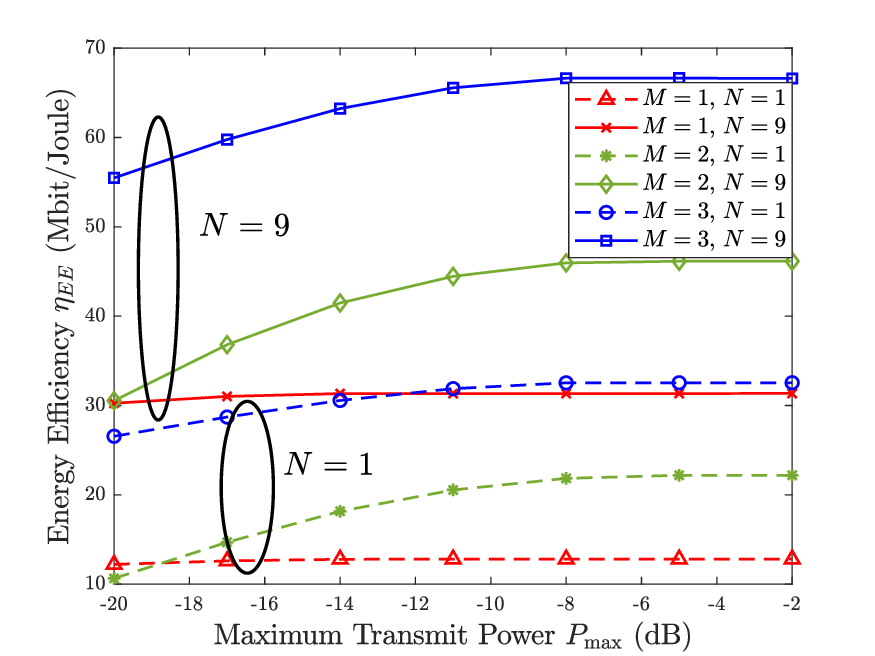}
		\caption{Energy efficiency in \text{Algorithm 2} under different number of waveguides and PAs, for $\varrho=200$ and $c_{\varrho}=1/0.5$.}
		\label{fig_EE_MN}
	\end{figure}
	
	Fig.~\ref{fig_EE_MN} presents the energy efficiency of the proposed algorithm applied to PASS across $M$ and $N$, exhibiting a markedly different trend and performance gain from Fig.~\ref{fig_EE_simulation}. For instance, under $(M, N,P_{\max})=(3,9,-2 \ \text{dB})$, the proposed algorithm achieves a $242.7\%$ increase in energy efficiency compared with the baseline. Moreover, the curves in Fig.~\ref{fig_EE_MN} do not exhibit the decreasing trend with increasing $N$ and $P_{\max}$ observed in Fig.~\ref{fig_EE_simulation}. This benefit arises from the carefully designed transmit precoding and PA radiation power allocation in \textbf{Algorithm 2}, which can enhance system performance under a limited power budget.

	\subsection{Impact of Tuning Grid Resolution}
	\begin{figure}[t]
		\centering
		\includegraphics[width=0.5\textwidth]{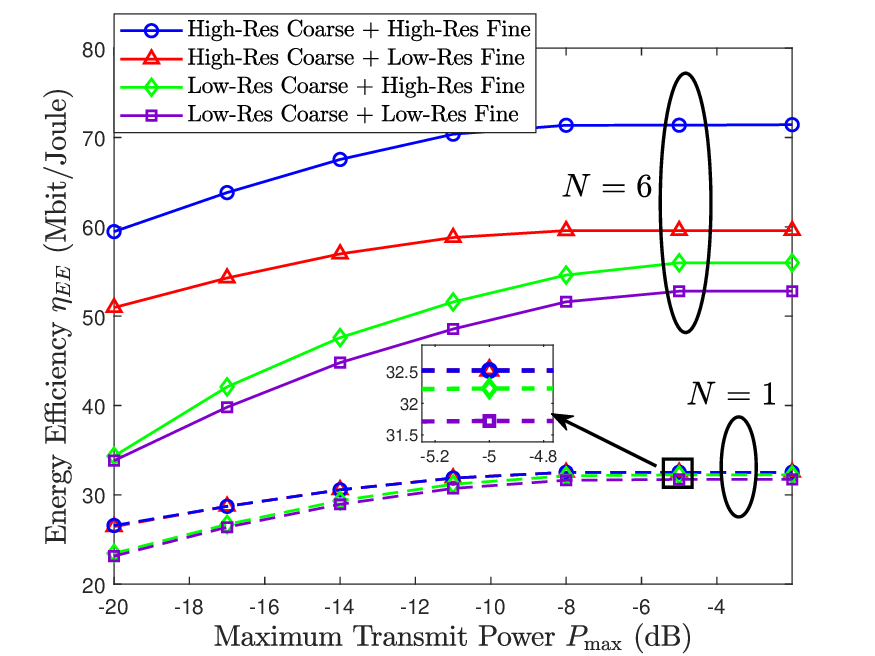}
		\caption{Comparison under high and low resolution for coarse-sliding and fine-tuning grids, for $\varrho=90$ and $c_{\varrho}=1/0.6$.}
		\label{fig_EE_grids}
	\end{figure}
	
	Fig.~\ref{fig_EE_grids} compares the energy efficiency of coarse-sliding under high-resolution $\delta_C=1$ m and low-resolution $\delta_C=10$ m, as well as fine-tuning under high-resolution $\delta_F=10^{-4}$ m and low-resolution $\delta_F=10^{-2}$ m. It can be observed in the single-PA scenario, the resolution of coarse-sliding and fine-tuning has a minor impact on communication performance. However, when $N=6$, the resolution of PA position deployment has a pronounced effect on the overall energy efficiency. This is because the $N$ PAs induce multi-path effects for each user, and inaccurate PA positions lead to phase misalignment among paths, resulting in substantial performance loss. Moreover, it shows that high-resolution in coarse-sliding is more critical than in fine-tuning, offering important guidance for selecting the motorized- and piezoelectric-based resolutions in the STT protocol.
	
	\subsection{Energy Efficiency Performance of the Proposed Protocols}
	\begin{figure}[t]
		\centering
		\includegraphics[width=0.5\textwidth]{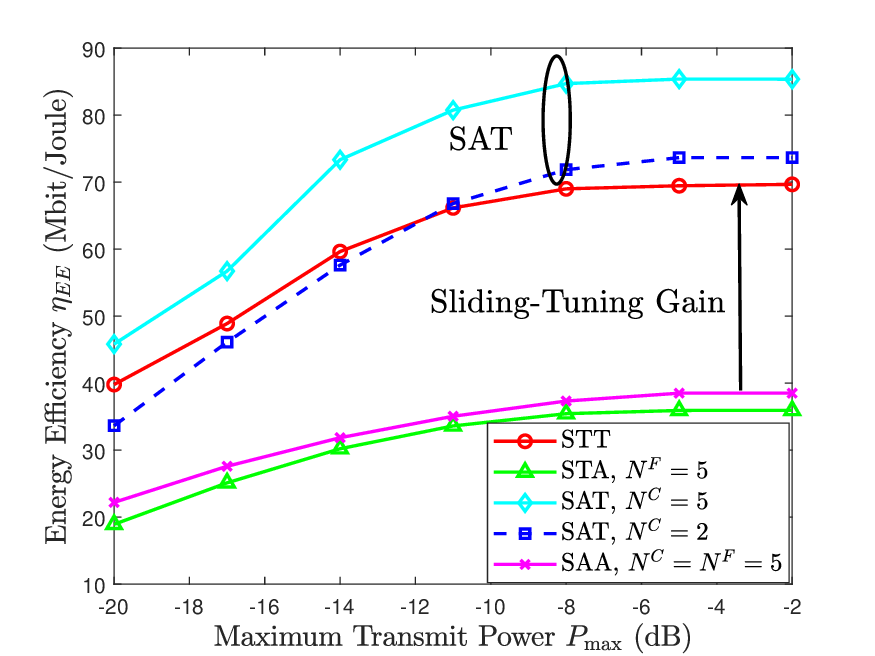}
		\caption{Comparison under different protocols, for $\varrho=450$ and $c_{\varrho}=1/0.6$.}
		\label{fig_protocol}
	\end{figure}
	Fig.~\ref{fig_protocol} demonstrates the energy efficiency of the proposed four protocols under varying $P_{\max}$. It is surprising that, although STT has the maximum DoF in PA position adjustment and is expected to deliver the best communication performance, it does not achieve the highest energy efficiency. Instead, SAT achieves the highest energy efficiency by eliminating the substantial motorized-module power consumption incurred during base sliding, which makes it an energy-efficient protocol worthy of further investigation within PASS. Notably, the SAT protocol can achieve comparable energy efficiency to STT by pre-deploying only twice as many bases, i.e., $N_C=2$. In contrast, the STA protocol incurs the base-sliding energy cost and simultaneously delivers unsatisfactory spectral efficiency, leading to the lowest energy efficiency. Compared with the SAA protocol, the STT protocol attains $80.8\%$ improvement in energy efficiency, evidencing a clear performance gain from coarse-sliding and fine-tuning.
	
	\section{Conclusion}\label{sec:conclusion}
	This paper has provided four implementation protocols in the DSD framework and proposed a novel low-complexity energy efficiency maximization algorithm, serving as a valuable reference for future PASS research and implementation. The proposed algorithm integrated Lagrangian transformations, block coordinate manifold optimization, and penalty-based alternating optimization to enable joint PASS design under each implementation protocol. Numerical results demonstrated substantial system gains with the proposed algorithm in PASS, and SAT was identified as the recommended protocol owing to its highest energy efficiency. 
	
	The proposed DSD framework for energy-efficient PA deployment could motivate future research on PASS, such as protocol design and channel estimation. More specifically, designing an appropriate transmission protocol in the DSD framework for PASS is a promising direction for achieving an optimal trade-off among system performance, power consumption, and response time. It could also explicitly account for dual-scale position adjustment speeds together with practical hardware constraints. In addition, dual-scale channel estimation and codebook design call for further investigation to accurately acquire CSI for this hybrid architecture.

	\appendices
	\section{Proof for Lemma \ref{lemma1}}\label{prof_lemma1}
	The derivation is based on the works in \cite{FP,WMMSE}, which equivalently transform the objective function by introducing auxiliary variables. We first use the Dinkelbach transform \cite{DT} to reformulate the objective function~\eqref{P1_EE} into
	\begin{equation}\label{SE2}
		\max_{\mathbf{W},\bm{\Xi},\mathbf{X}} B \cdot \sum_{k=1}^K R_k - q\cdot P_{all},
	\end{equation}
	with a new auxiliary variable $q$ and iteratively updated by $q = B \cdot\sum_{k=1}^K R_k/ P_{\text{all}}$. Then, we construct a new optimization problem and prove its equivalence to the problem~\eqref{SE2} as
	\begin{subequations}\label{error2}
		\begin{align}
			\min_{\mathbf{W},\bm{\Xi},\mathbf{X}, \bm{\kappa}, \mathbf{t}} \quad & B \cdot \sum_{k=1}^K \left(  \kappa_k \epsilon_k - \log(\kappa_k)  \right) + q\cdot P_{all} \\
			\mathrm{s.t.} \quad & \eqref{P1_1},\ \eqref{P1_2},\ \eqref{P1_3},
		\end{align}
	\end{subequations}
	where $\bm{\epsilon}=[\epsilon_1,\cdots,\epsilon_K]^T$ is given by~\eqref{epsilon}. For fixed $\mathbf{W}$, $\bm{\Xi}$, and $\mathbf{X}$, the objective function of \eqref{error2} is strictly convex with respect to $\kappa_k$ and $t_k$ for each $k$. Therefore, the optimal $\bm{\kappa}$ and $\mathbf{t}$ can be obtained by setting the corresponding first-order conditions to zero, yielding $\kappa_k=\epsilon_k^{-1}$ and 
	\begin{equation}
		t_k \!=\! \frac{ \mathbf{\bar{h}}_k^T \bm{\Xi} \mathbf{G}  \mathbf{w}_k }{ \sum_{i=1}^K\! \left(  \left| \mathbf{\bar{h}}_k^T \bm{\Xi} \mathbf{G} \mathbf{w}_i \right|^2 \!\!+\!\! \frac{C_0}{K_r+1}  \mathbf{w}_i^T  \mathbf{G}^T \bm{\Xi}^T \mathbf{R}_k \bm{\Xi} \mathbf{G}^* \mathbf{w}_i^* \right)  \!\!+\!\! N_0}\!.
	\end{equation}

	Then, we substitute the optimal $\bm{\kappa}$ and $\mathbf{t}$ into the objective function of problem~\eqref{error2} and obtain
	\begin{equation}
		\min_{\mathbf{W},\bm{\Xi},\mathbf{X}} B \cdot \sum_{k=1}^K (1-\log_2(\epsilon_k^{-1})) + q \cdot P_{all},
	\end{equation}
	which can be proven to be equivalent to~\eqref{SE2}. Hence, for fixed optimal $\bm{\kappa}$ and $\mathbf{t}$, and dropping irrelevant constant terms w.r.t. $\{\mathbf{W},\bm{\Xi},\mathbf{X}\}$ in the objective function of \eqref{error2}, the problem~\eqref{error2} reduces to problem~\eqref{error1}.


\begin{thebibliography}{11}
		\bibitem{6G}
		ITU-R. \emph{Framework and Overall Objectives of the Future Development of IMT for 2030 and Beyond,} Recommendation ITU-R M.2160-0, Nov. 2023. [Online]. Available: https://www.itu.int/rec/R-REC-M.2160-0-202311-I/en
		
		\bibitem{RIS}
		C. Huang, A. Zappone, G. C. Alexandropoulos, M. Debbah and C. Yuen, ``Reconfigurable intelligent surfaces for energy efficiency in wireless communication,'' \emph{IEEE Trans. Wireless Commun.}, vol. 18, no. 8, pp. 4157-4170, Aug. 2019.
		

		\bibitem{fluid1}
		T.~Wu, K.~Zhi, J.~Yao, X.~Lai, J.~Zheng, H.~Niu, M.~Elkashlan, K.-K.~Wong, C.-B.~Chae, Z.~Ding, G.~K.~Karagiannidis, M.~Debbah, and C.~Yuen, ``Fluid antenna systems enabling {6G}: Principles, applications, and research directions,'' \emph{IEEE Wireless Commun.}, early access, Dec. 2025, doi: 10.1109/MWC.2025.3629597.
		
		
		\bibitem{movable1}
		L.~Zhu, W.~Ma, W.~Mei, Y.~Zeng, Q.~Wu, B.~Ning, Z.~Xiao, X.~Shao, J.~Zhang, and R.~Zhang, ``A tutorial on movable antennas for wireless networks,'' \emph{IEEE Commun. Surv. Tutor.}, vol. 28, pp. 3002-3054, Feb. 2025.
		
		\bibitem{PA_mag}
		Y. Liu, Z. Wang, X. Mu, C. Ouyang, X. Xu and Z. Ding, ``Pinching-antenna systems: Architecture designs, opportunities, and outlook,'' \emph{IEEE Commun. Mag.}, vol. 64, no. 1, pp. 190-196, Jan. 2026.
		
		
		\bibitem{PA_ding}
		Z. Ding, R. Schober and H. Vincent Poor, ``Flexible-antenna systems: A pinching-antenna perspective,'' \emph{IEEE Trans. Commun.}, vol. 73, no. 10, pp. 9236-9253, Oct. 2025.
		
		\bibitem{tutorial}
		Y. Liu, H. Jiang, X. Xu, Z. Wang, J. Guo, C. Ouyang, X. Mu, Z. Ding, A. Nallanathan, G. K. Karagiannidis, R. Schober, ``Pinching-antenna systems (PASS): A tutorial'', \emph{IEEE Trans. Commun.}, vol. 74, pp. 4881-4918, Jan. 2026.
		
		\bibitem{LoS_links}
		K. Wang, C. Ouyang, Y. Liu and Z. Ding, ``Pinching-antenna systems with LoS blockages,'' \emph{IEEE Wireless Commun. Lett.}, vol. 14, no. 12, pp. 4122-4126, Dec. 2025.
		
		
		\bibitem{security1}
		M. Sun, C. Ouyang, S. Wu, Y. Liu, ``Physical layer security for pinching-antenna systems (PASS),'' \emph{IEEE Trans. Wireless Commun.}, early access, 2025.
		
		\bibitem{security2}
		K. Wang, Z. Ding, N. Al-Dhahir, ``Pinching-antenna systems for physical layer security,'' \emph{IEEE Wireless Commun. Lett.}, vol. 15, pp. 260-264, 2026.
		
		
		
		\bibitem{docomo}
		A. Fukuda, H. Yamamoto, H. Okazaki, Y. Suzuki, and K. Kawai,
		``Pinching antenna - using a dielectric waveguide as an antenna,'' \emph{NTT DOCOMO Technical J.}, vol. 23, no. 3, pp. 5-12, Jan. 2022.
		
		
		\bibitem{model_PA}
		Z. Wang, C. Ouyang, X. Mu, Y. Liu, Z. Ding, ``Modeling and beamforming optimization for pinching-antenna systems'', \emph{IEEE Trans. Commun.}, vol. 73, no. 12, pp. 13904-13919, Dec. 2025.
		
		
		
		\bibitem{PA_equal1}
		S. Shan, C. Ouyang, Y. Li, Y. Liu, ``Exploiting pinching-antenna systems in multicast communications,'' \emph{IEEE Trans. Commun.}, vol. 74, pp. 419-432, 2026.

		
		\bibitem{PA_general}
		Y. Xu, D. Xu, X. Yu, S. Song, Z. Ding, R. Schober, ``Joint radiation power, antenna position, and beamforming optimization for pinching-antenna systems with motion power consumption,'' \emph{IEEE Trans. Wireless Commun.}, vol. 25, pp. 7825-7841, 2026.
		
		\bibitem{PA_discrete1}
		X. Xu, X. Mu, Z. Wang, Y. Liu, A. Nallanathan, ``Pinching-antenna systems (PASS): Power radiation model and optimal beamforming design,'' \emph{IEEE Trans. Commun.}, vol. 74, pp. 2160-2175, 2026.
		
		\bibitem{PA_discrete2}
		B. Zhang, H. Zhang, K. Yang, Y. Zhao, K. Wang, ``On the performance analysis of pinching-antenna-enabled SWIPT systems,'' \emph{IEEE Trans. Veh. Technol.}, early access, 2025.
		

		\bibitem{two_stage1}
		Y. Xu, Z. Ding and G. K. Karagiannidis, ``Rate maximization for downlink pinching-antenna systems,'' \emph{IEEE Wireless Commun. Lett.}, vol. 14, no. 5, pp. 1431-1435, May 2025.
		
		
		\bibitem{two_stage2}
		Z. Ding and H. V. Poor, ``Analytical optimization for antenna placement in pinching-antenna systems,'' \emph{arXiv preprint arXiv:2507.13307}, 2025.
		
		\bibitem{two_stage3}
		X. Xie, F. Fang, Z. Ding and X. Wang, ``A low-complexity placement design of pinching-antenna systems," \emph{IEEE Commun. Lett.}, vol. 29, no. 8, pp. 1784-1788, Aug. 2025.
		
		
		\bibitem{EM_field}
		D. K. Cheng, \textit{Field and wave electromagnetics}, Addison-Wesley, 1989.
		
		\bibitem{PI}
		Physik Instrumente (PI) GmbH \& Co. KG, \emph{Miniature Precision Positioning Stages}. PI Catalog, 2015. [Online]. Available: \url{https://www.pi-usa.us/fileadmin/user_upload/pi_us/files/catalogs/PI_Precision_Miniature_Translation_Stages.pdf}
		
		
		\bibitem{NR_frame}
		S. Parkvall, E. Dahlman, A. Furuskar and M. Frenne, ``NR: The new 5G radio access technology,'' \emph{IEEE Commun. Stand. Mag.}, vol. 1, no. 4, pp. 24-30, Dec. 2017.
		
		\bibitem{cellfree_model}
		H. Q. Ngo, A. Ashikhmin, H. Yang, E. G. Larsson and T. L. Marzetta, ``Cell-free massive MIMO versus small cells,'' \emph{IEEE Trans. Wireless Commun.}, vol. 16, no. 3, pp. 1834-1850, Mar. 2017.
		
		
		\bibitem{Riemannian}
		P.-A. Absil, R. Mahony, and R. Sepulchre, \emph{Optimization Algorithms on Matrix Manifolds.} Princeton, NJ, USA: Princeton Univ. Press, 2009.
		
		\bibitem{step_size}
		S. Hosseini, W. Huang, and R. Yousefpour, ``Line search algorithms for locally Lipschitz functions on Riemannian manifolds,'' \emph{SIAM J. Optim.}, vol. 28, no. 1, pp. 596-619, 2018.
		
		\bibitem{pdd}
		Q. Shi and M. Hong, ``Penalty dual decomposition method for nonsmooth nonconvex optimization—Part I: Algorithms and convergence analysis,'' \emph{IEEE Trans. Signal Process.}, vol. 68, pp. 4108-4122, Jun. 2020.
		
		\bibitem{mrt}
		X. Gan, C. Zhong, C. Huang, Z. Yang and Z. Zhang, ``Multiple RISs assisted cell-free networks with two-timescale CSI: Performance analysis and system design,'' \emph{IEEE Trans. Commun.}, vol. 70, no. 11, pp. 7696-7710, Nov. 2022.
		
		\bibitem{FP}
		K. Shen and W. Yu, ``Fractional programming for communication systems—Part I: Power control and beamforming,'' \emph{IEEE Trans. Signal Process.}, vol. 66, no. 10, pp. 2616-2630, May 2018.
		
		\bibitem{WMMSE}
		Q. Shi, M. Razaviyayn, Z. -Q. Luo and C. He, ``An iteratively weighted MMSE approach to distributed sum-utility maximization for a MIMO interfering broadcast channel,'' \emph{IEEE Trans. Signal Process.}, vol. 59, no. 9, pp. 4331-4340, Sept. 2011.
		
		\bibitem{DT}
		W. Dinkelbach, ``On nonlinear fractional programming,'' \emph{Manage. Sci.}, vol. 13, no. 7, pp. 492-498, Mar. 1967.
		
		
	\end{thebibliography}
\end{document}